\documentclass[aps,prb,reprint,superscriptaddress]{revtex4-1}

\usepackage{graphicx}
\usepackage{dcolumn}
\usepackage{bm}
\usepackage{amsmath}
\usepackage{amssymb}
\usepackage{hyperref}
\usepackage{xcolor}
\usepackage{gensymb}
\usepackage{pdfpages}
\usepackage{pgffor}

\makeatletter
\AtBeginDocument{\let\LS@rot\@undefined}
\makeatother

\begin{document}


\title{Absence of a giant spin Hall effect in plasma-hydrogenated graphene}

\author{Tobias V\"olkl}
\affiliation{Institut f\"ur Experimentelle und Angewandte Physik, Universit\"at Regensburg, Germany}
\author{Denis Kochan}
\affiliation{Institut f\"ur Theoretische Physik, Universit\"at Regensburg, Germany}
\author{Thomas Ebnet}
\affiliation{Institut f\"ur Experimentelle und Angewandte Physik, Universit\"at Regensburg, Germany}
\author{Sebastian Ringer}
\affiliation{Institut f\"ur Experimentelle und Angewandte Physik, Universit\"at Regensburg, Germany}
\author{Daniel Schiermeier}
\affiliation{Institut f\"ur Experimentelle und Angewandte Physik, Universit\"at Regensburg, Germany}
\author{Philipp Nagler}
\affiliation{Institut f\"ur Experimentelle und Angewandte Physik, Universit\"at Regensburg, Germany}
\author{Tobias Korn}
\affiliation{Institut f\"ur Experimentelle und Angewandte Physik, Universit\"at Regensburg, Germany}
\author{Christian Sch\"uller}
\affiliation{Institut f\"ur Experimentelle und Angewandte Physik, Universit\"at Regensburg, Germany}
\author{Jaroslav Fabian}
\affiliation{Institut f\"ur Theoretische Physik, Universit\"at Regensburg, Germany}
\author{Dieter Weiss}
\affiliation{Institut f\"ur Experimentelle und Angewandte Physik, Universit\"at Regensburg, Germany}
\author{Jonathan Eroms}
\email{jonathan.eroms@ur.de}
\affiliation{Institut f\"ur Experimentelle und Angewandte Physik, Universit\"at Regensburg, Germany}





\date{\today}

\begin{abstract}

The weak spin-orbit interaction in graphene was predicted to be increased, {\em e.g.}, by hydrogenation. This should result in a sizable spin Hall effect (SHE). We employ two different methods to examine the spin Hall effect in weakly hydrogenated graphene. For hydrogenation we expose graphene to a hydrogen plasma and use Raman spectroscopy to characterize this method. We then investigate the SHE of hydrogenated graphene in the H-bar method and by direct measurements of the inverse SHE. Although a large nonlocal resistance can be observed in the H-bar structure, comparison with the results of the other method indicate that this nonlocal resistance is caused by a non-spin-related origin.
\end{abstract}	

\maketitle

\section{Introduction}
Covalently bonded hydrogen was predicted to significantly increase the spin-orbit coupling (SOC) of graphene by Castro Neto and Guinea\cite{CastroNeto2009}. However, experimental results on this were conflicting. Balakrishnan~\emph{et al.}~ reported a high nonlocal resistance in weakly hydrogenated graphene in the so called H-bar structure\cite{Balakrishnan2013a}. They further observed an oscillatory behavior of this nonlocal resistance with an in-plane magnetic field and therefore attributed this effect to the SHE with a spin Hall angle of around $\alpha_{SH} = 0.18 - 0.45$. A high nonlocal resistance in similar samples was also observed by Kaverzin and van Wees\cite{Kaverzin2015}. However they obtained an unrealistically high value for the spin Hall angle of $\alpha_{SH} =1.5$ and could not observe any effect of an in-plane magnetic field on this nonlocal resistance. They therefore argue that this nonlocal signal has a non spin related origin.

Here, we perform different types of experiments to solve this controversy. For hydrogenation we expose graphene to a hydrogen plasma which has several advantages over the hydrogenation method by exposing hydrogen silsesquioxane (HSQ) to an electron beam, employed in Refs. \onlinecite{Balakrishnan2013a,Kaverzin2015}. We use Raman spectroscopy to characterize graphene exposed to hydrogen or deuterium to verify that the created defects by this method are indeed bonded hydrogen atoms. Then we perform non-local measurements in the so-called H-bar geometry in graphene that was hydrogenated by this method. Further, we employ electrical spin injection into hydrogenated graphene to perform spin transport measurements as well as measurements of the inverse spin Hall effect. Our results show that the large nonlocal signal in hydrogenated graphene is not related to the spin Hall effect.

\section{Plasma hydrogenation of graphene}

Due to limitations of the HSQ-based hydrogenation procedure, which we describe in more detail below, 
we explore hydrogenation by exposing graphene to a hydrogen plasma in a reactive ion etching chamber (RIE). Following the recipe developed by Wojtaszek~\emph{et al.}\cite{Wojtaszek2011}, exfoliated graphene was exposed to hydrogen plasma of pressure $p=40$ mTorr, 30 sccm gas flow and 2 W power. The relatively low power leads to a low acceleration bias voltage of $U_{bias}<2$ V, which reduces the creation of lattice defects. 
The samples were then investigated by Raman spectroscopy.

\begin{figure*}
	\centering
	\hspace{3mm}\includegraphics[width=0.6538\columnwidth]{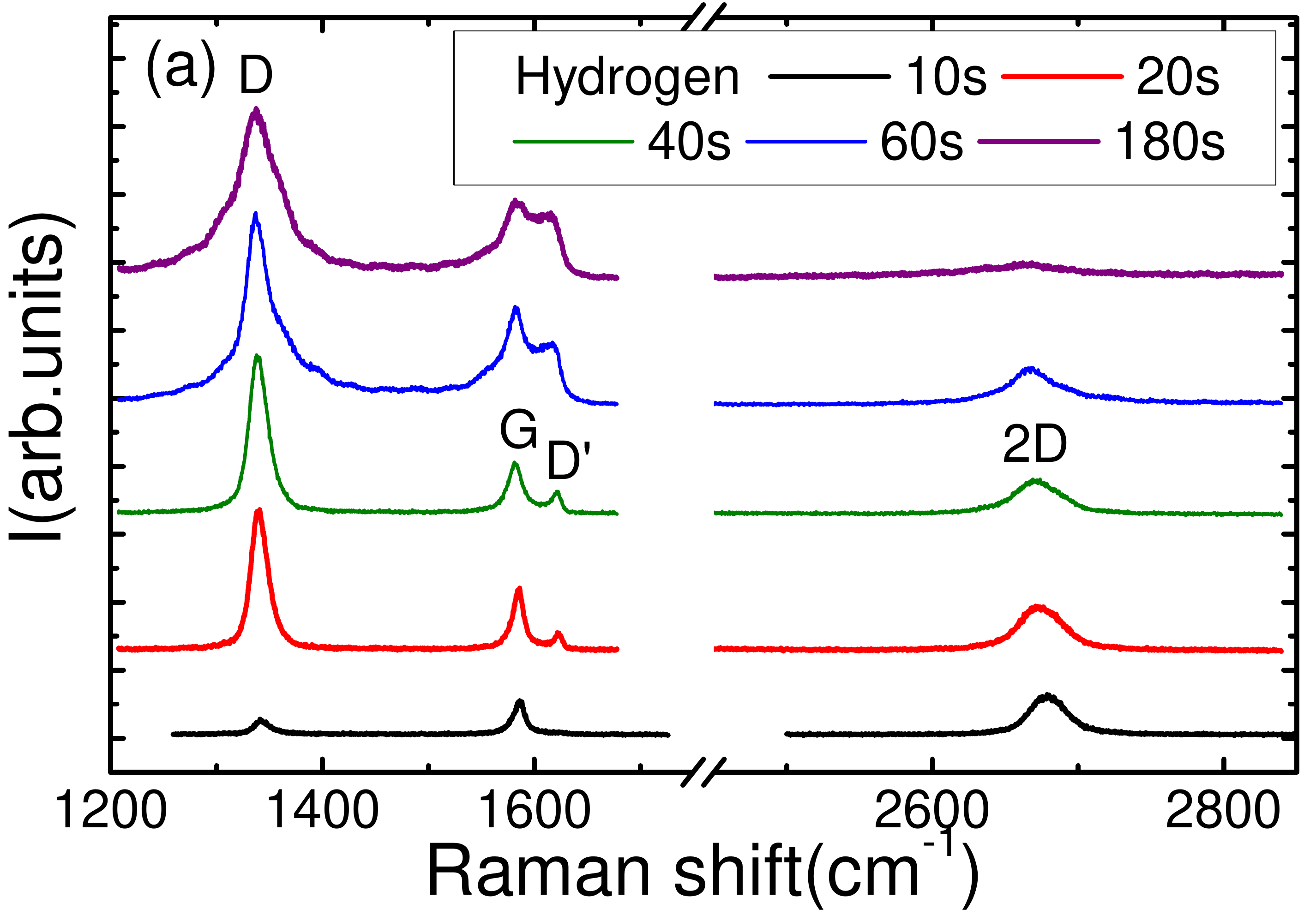}%
	\hspace{0.89cm}%
	\includegraphics[width=0.6599\columnwidth]{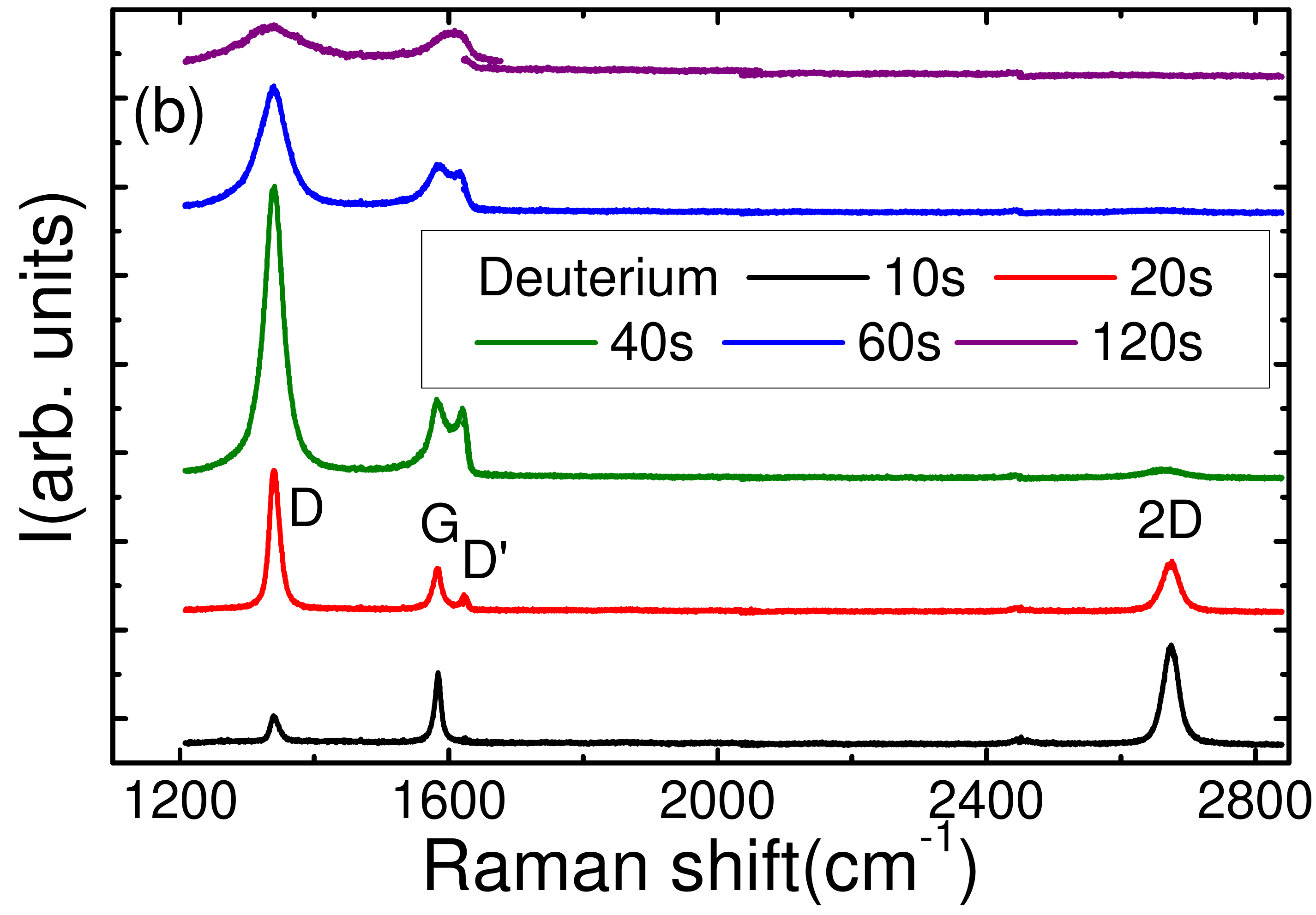}
	\includegraphics[width=0.6919\columnwidth]{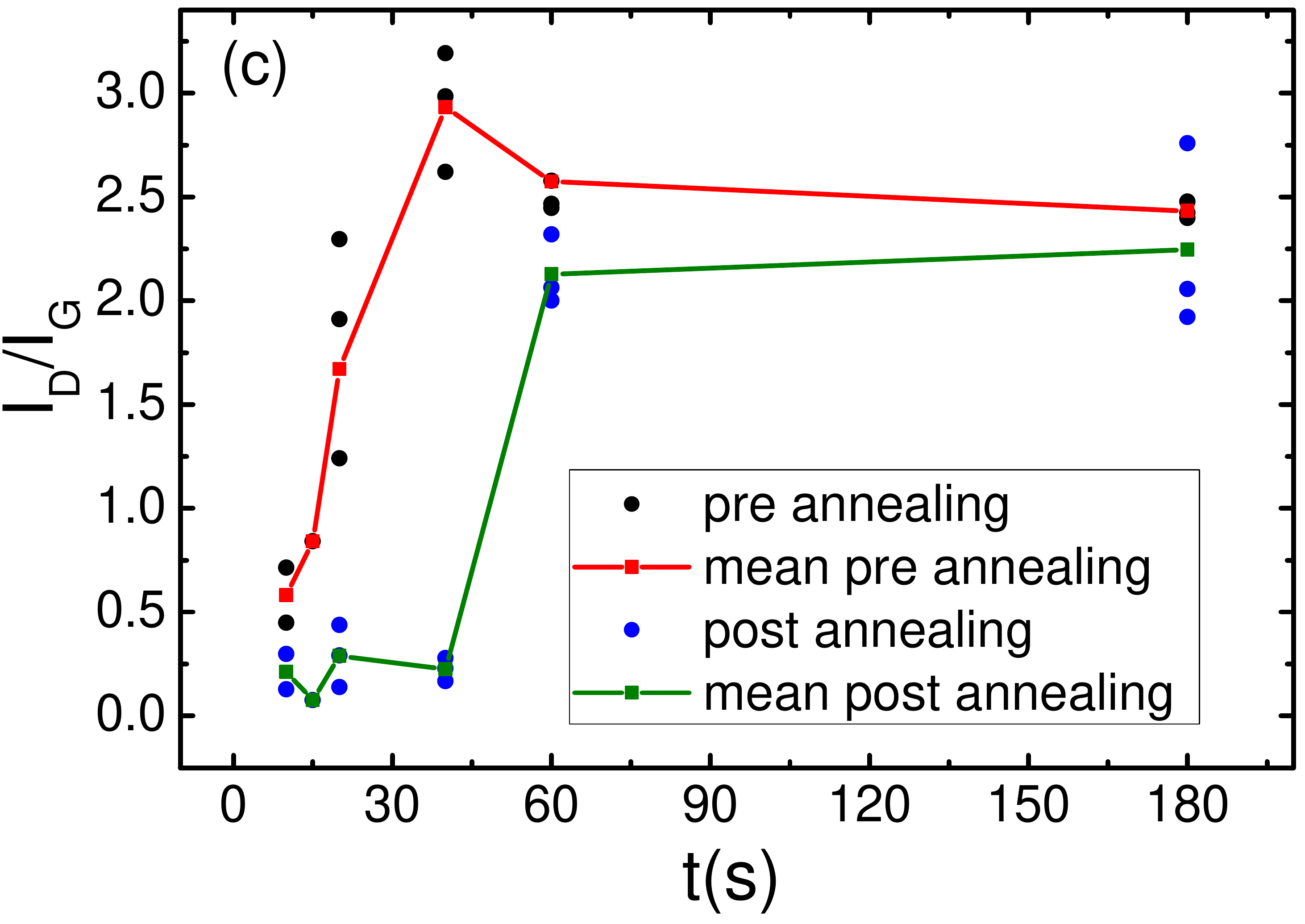}%
	\hspace{0.5cm}%
	\includegraphics[width=0.7\columnwidth]{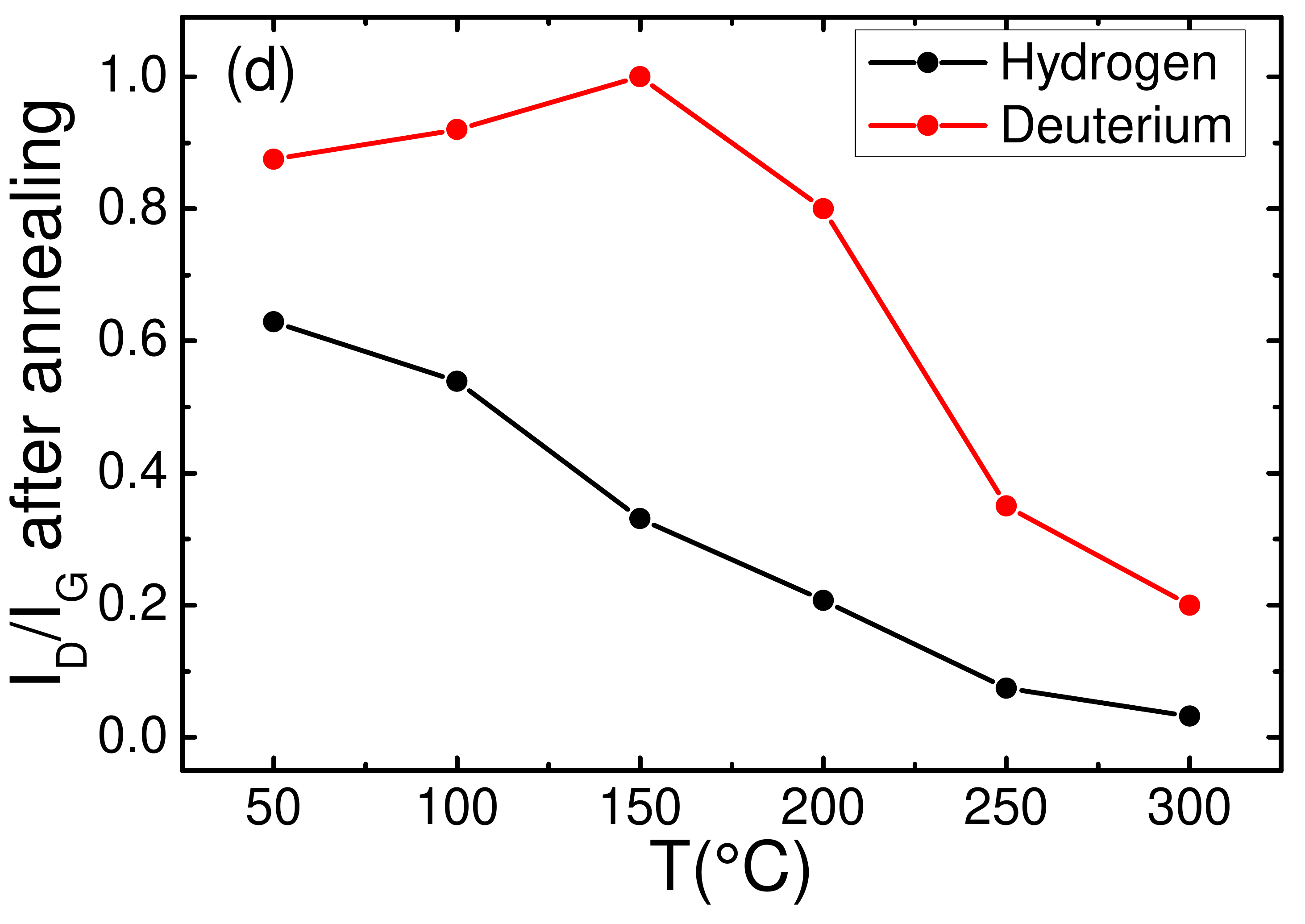}%
	\caption[Raman spectra for different exposure times with hydrogen plasma]{(a) Raman spectra for different exposure times with hydrogen plasma. An increase of D and D$^\prime$-peak intensities as well as a decrease of the 2D-peak intensity with increasing plasma exposure time can be observed, indicating the creation of defects. (b) Raman spectra for different exposure times with deuterium plasma. The deuterium seems to create more defects than hydrogen for the same exposure times. (c) Ratio between D and G-peak intensities with hydrogen plasma exposure time before (red curve) and after (green curve) annealing at $T=320 \degree$C. $I_D/I_G$ increases up to a exposure time of $t=40$ s and decreases for high exposure times. For low exposure times the hydrogenation process is reversible. (d) $I_D/I_G$ after annealing at different temperatures normalized to its initial value for hydrogen (black dots) and deuterium (red dots) with a plasma time of $t=20$ s. Deuterium is more stable with increasing temperature than hydrogen. This is a strong indication that the defects created by this method are bonded hydrogen (deuterium) atoms}
	\label{hydrogen_dose}
\end{figure*}

Fig. \ref{hydrogen_dose} (a) shows Raman spectra of samples with different plasma exposure time. 
With increasing exposure time both a D-peak and a D$^\prime$-peak arise, which indicate the presence of defects. For higher exposure times a decrease of the 2D-peak intensity can be observed which indicates an alteration of the electronic band structure. As can be seen in the red curve in Fig. \ref{hydrogen_dose} (c) the ratio between the D and G-peak intensities increases with exposure time up to a value around $I_D/I_G=3$ for an exposure time of $t=40$ s and decreases for longer exposure times. 
For low defect densities the ratio between D and G-peak intensities is proportional to the defect density\cite{Cancado2011}:
\begin{equation}
	n_D(cm^{-2})=\frac{1.8 \pm 0.5 \cdot 10^{22}}{\lambda_L^4} \left( \frac{I_D}{I_G} \right) 
	\label{n_D}
\end{equation}
with $\lambda_L=532$ nm (given in nm in Eq. \eqref{n_D}) being the excitation wavelength. $I_D/I_G$ reaches its maximum when the average distance between defects becomes comparable to the distance an e-h pair travels in its lifetime, given by $l_x=v_F/\omega_D$ with $\omega_D$ being the D-peak frequency\cite{Cancado2011}. At higher defect densities the D-peak becomes broader and its intensity decreases. Further, at high defect densities the graphene band structure is altered by the defects, which reduces possible transitions\cite{Ferrari2007}. Since the 2D peak is double resonant it is more sensitive to this alteration than the D- and G-peaks and therefore a reduction of the 2D-peak intensity with increasing exposure time can be observed in Fig. \ref{hydrogen_dose}(a).

The green curve of Fig. \ref{hydrogen_dose}(c) shows $I_D/I_G$ for the same samples after annealing in vacuum at 320 $\degree$C for 1 h. For low plasma exposure times $t \leq 40$ s annealing almost fully removes the defects. Since this temperature is too low to heal vacancies\cite{Zion2017} in graphene, this behavior indicates that for these low exposure times the observed defects are bonded hydrogen atoms. For $t>40$ s the defects could not be removed by annealing. Therefore the occurrence of lattice defects for higher plasma exposure times is likely. Possible explanations for this might be heating of the samples during the exposure process or etching of carbon atoms by the formation of CH$_2$ after saturation of the hydrogen coverage of graphene\cite{Luo2009}.

To further determine the type of the observed defects the same experiment was performed with deuterium instead of hydrogen. Fig. \ref{hydrogen_dose}(b) shows Raman spectra for different exposure times. In comparison to Fig. \ref{hydrogen_dose}(a) deuterium seems to induce slightly more defects than hydrogen as can be seen by the rapid decrease of 2D-peak intensity in Fig. \ref{hydrogen_dose}(b). One explanation for this could be a higher reactivity of deuterium, due to a slightly increased binding energy\cite{Paris2013}. Another explanation is that the deuterium atoms are more likely to create lattice defects due to their higher mass.

Samples exposed to either hydrogen or deuterium with an exposure time of $t=20$ s were annealed for 1 h in vacuum at different temperatures. Fig. \ref{hydrogen_dose}(d) shows the relative $I_D/I_G$ ratio divided by its value before annealing. Surprisingly the bonded deuterium (red dots in Fig. \ref{hydrogen_dose}(d)) is more stable with temperature than the hydrogen (black dots in Fig. \ref{hydrogen_dose}(d)). A similar behavior has been observed for hydrogen and deuterium on graphite\cite{Zecho2002}. This can be explained by a slightly increased binding energy of deuterium due to zero-point energy effects\cite{Paris2013} and a lower attempt frequency due to the higher mass of deuterium compared to hydrogen, hindering desorption \cite{Zecho2002}. The fact that a different desorption behavior was found for hydrogen and deuterium is a clear indication that the defects created by this method are really bonded hydrogen since there should be no difference for other defect types.

Concerning the HSQ-based hydrogenation method employed in Refs. \onlinecite{Balakrishnan2013a,Kaverzin2015,Ryu2008} we note several difficulties. 
First, the HSQ film cannot be removed after exposure without destroying the underlying graphene sheet. Therefore, hydrogenation can only be done as a last step of the sample fabrication. Since resist residues from previous steps proved to prevent efficient hydrogenation, it is expected that the hydrogen coverage produced by this method is not homogeneous. Second, a high p-type doping was always observed in samples produced by this method both in our measurements \footnote{see supplemental material} as well as in the measurements by Kaverzin and van Wees \cite{Kaverzin2015}. This is problematic since the occurrence of the SHE is only expected close to the charge neutrality point (CNP) \cite{Ferreira2014}, which in these samples is often not accessible due to the high doping. Third, it is not entirely clear that the defects produced by this method are really bonded hydrogen since the Raman measurements are not sensitive to the defect type. Therefore, in our experiments, we resort to plasma hydrogenation.

\section{Nonlocal resistance in hydrogenated graphene}
\begin{figure}
	\centering
	\includegraphics[width=0.9\columnwidth]{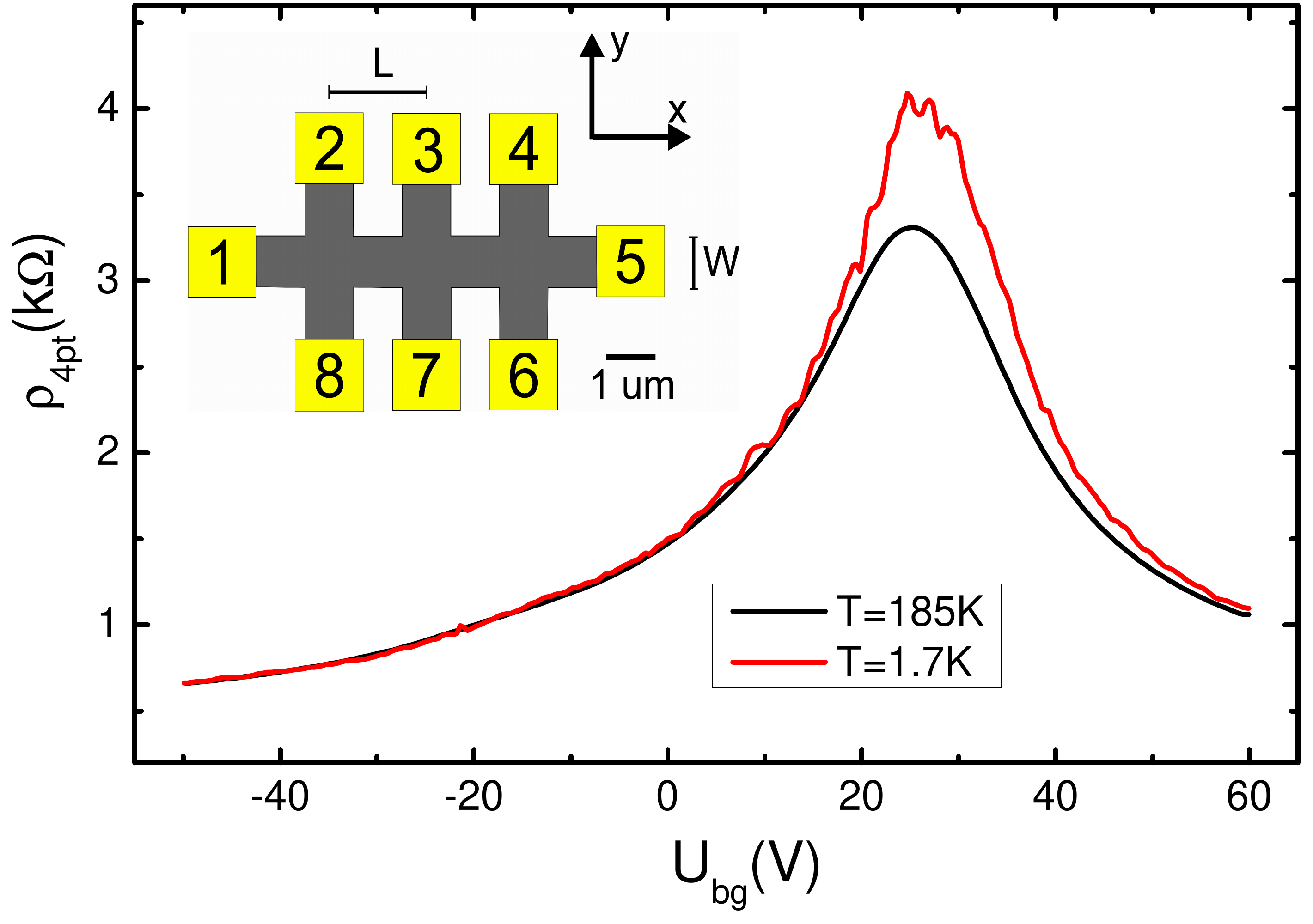}
	\caption[Optical microscope and schematic picture of a H-bar sample]{Back gate dependent four-point resistivity of H-bar sample at $T=185$ K (black curve) and $T=1.7$ K (red curve). This gives a position of the charge neutrality point at U$_{\textrm{CNP}}=26$ V indicating p-type doping and a mobility around $\mu=1600$ cm$^2$/Vs. Inset: Schematic picture of an H-bar sample.}
	\label{hstruct}
\end{figure}

Using plasma hydrogenation a Hall-bar sample was fabricated. First, exfoliated graphene was exposed to hydrogen plasma for 20 s as described in the previous section. Afterwards, oxygen plasma was used to etch the graphene into a Hall bar and 0.5 nm Cr + 60 nm Au were deposited for contacts. A schematic picture of the sample structure is displayed in the inset of Fig. \ref{hstruct}.
Raman measurements of this sample reveal $I_D/I_G=0.43$. Using Eq.~\ref{n_D} and assuming that the defect density equals the hydrogen atom density, we extract a coverage of 0.0025\%.
This value is much lower than in the previous section for the same exposure time since several lithography steps and therefore resist bake-out steps were necessary after the hydrogenation process. However, employing hydrogenation as a first step in the sample fabrication process was preferred over using it as a last step since it is expected that resist residues lead to an inhomogeneous hydrogen coverage of the sample.

Back gate sweeps of the 4-point resistivity of this sample at temperatures $T=185$ K (black curve) and $T=1.7$ K (red curve) are depicted in Fig. \ref{hstruct}. In this sample a p-type doping with $U_{CNP}=26$ V and mobilities of $\mu_h=1400$ cm$^2/$Vs /($\mu_h=1500$ cm$^2/$Vs) for the hole side and $\mu_{el}=1800$ cm$^2/$Vs ($\mu_{el}=2000$ cm$^2/$Vs) for the electron side at $T=185$ K ($T=1.7$ K) were observed.

\begin{figure*}
	\centering
	\includegraphics[width=0.65\columnwidth]{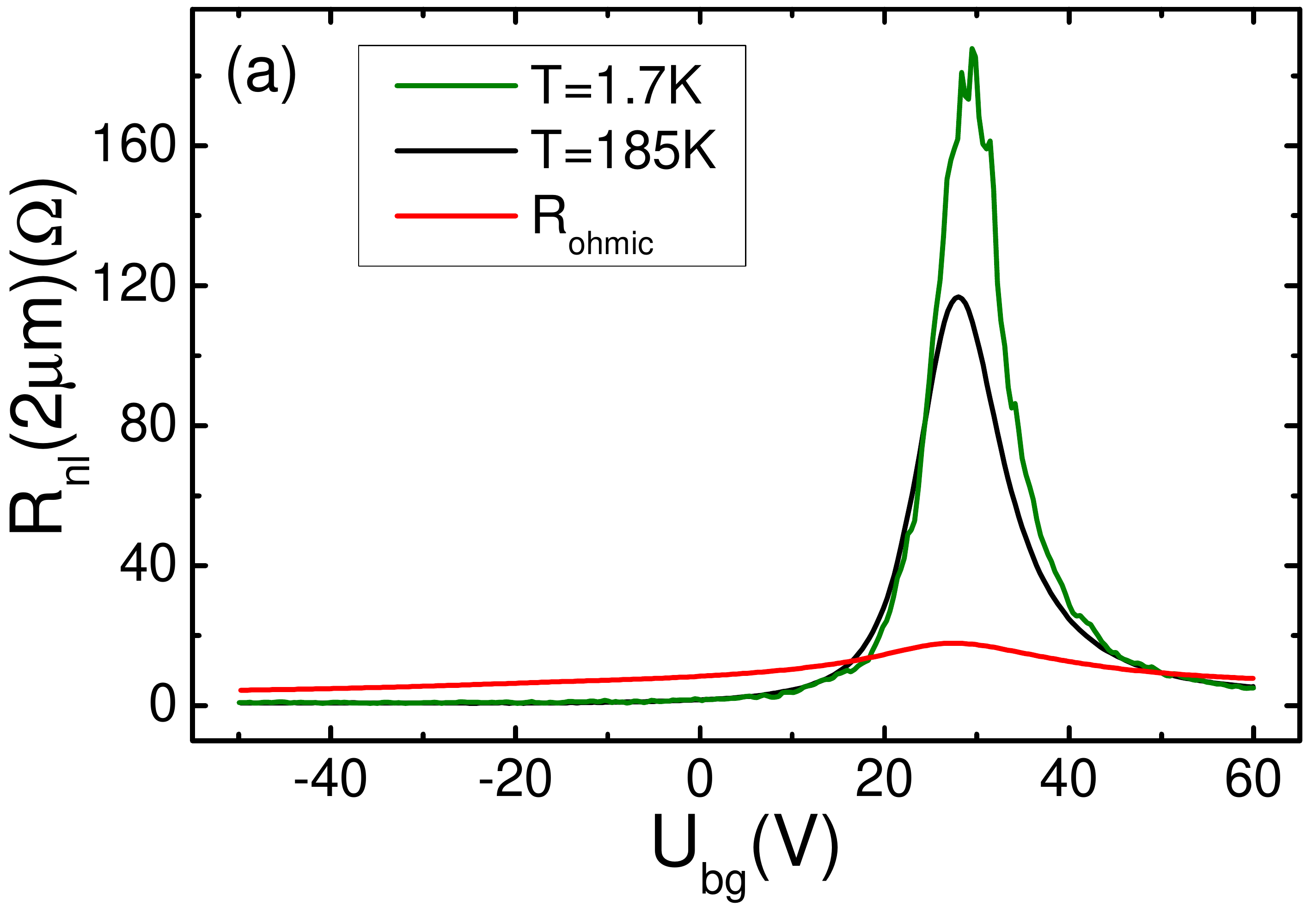}
	\hspace{0.2cm}
	\includegraphics[width=0.65\columnwidth]{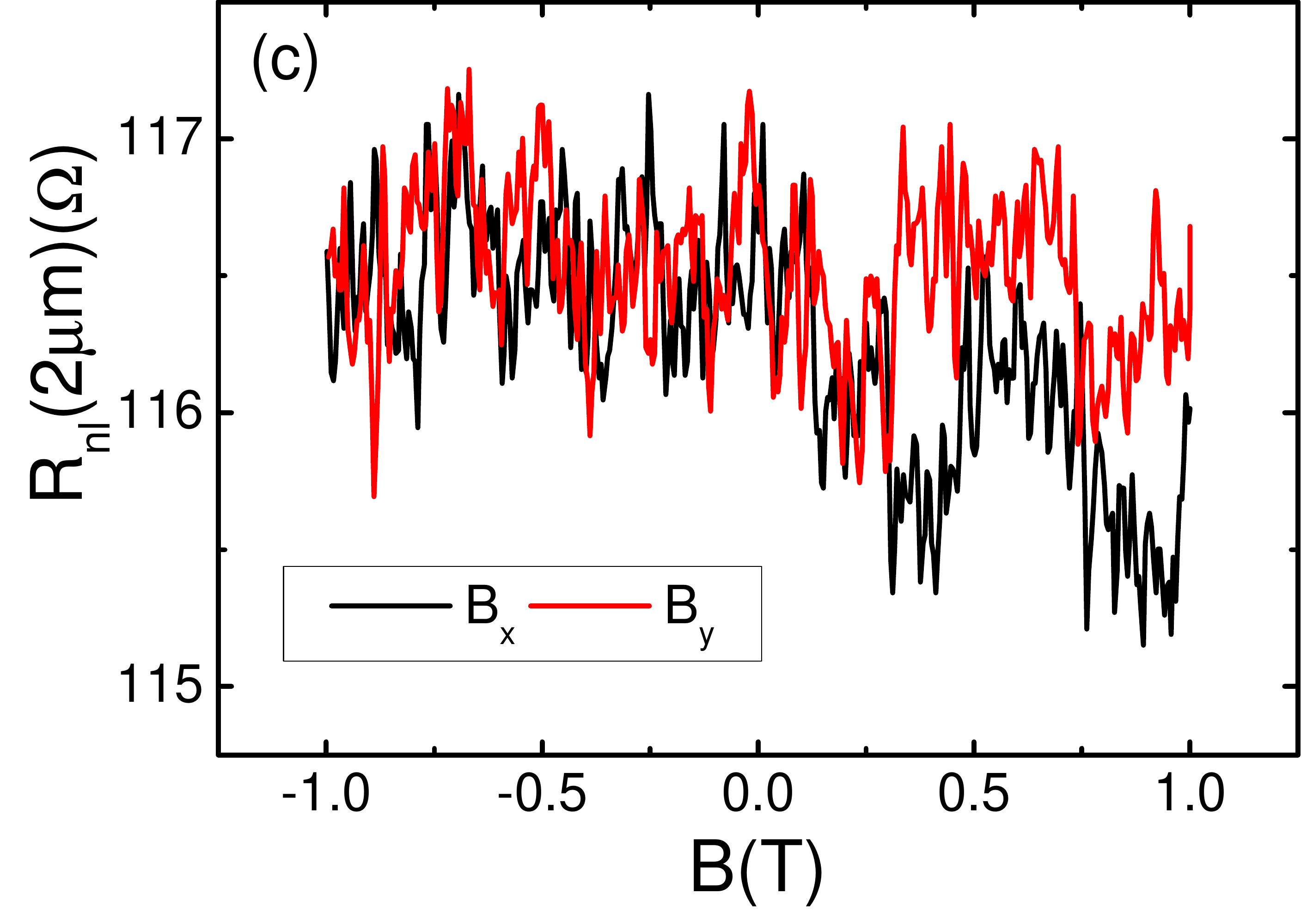}
	\hspace{0.2cm}
	\includegraphics[width=0.65\columnwidth]{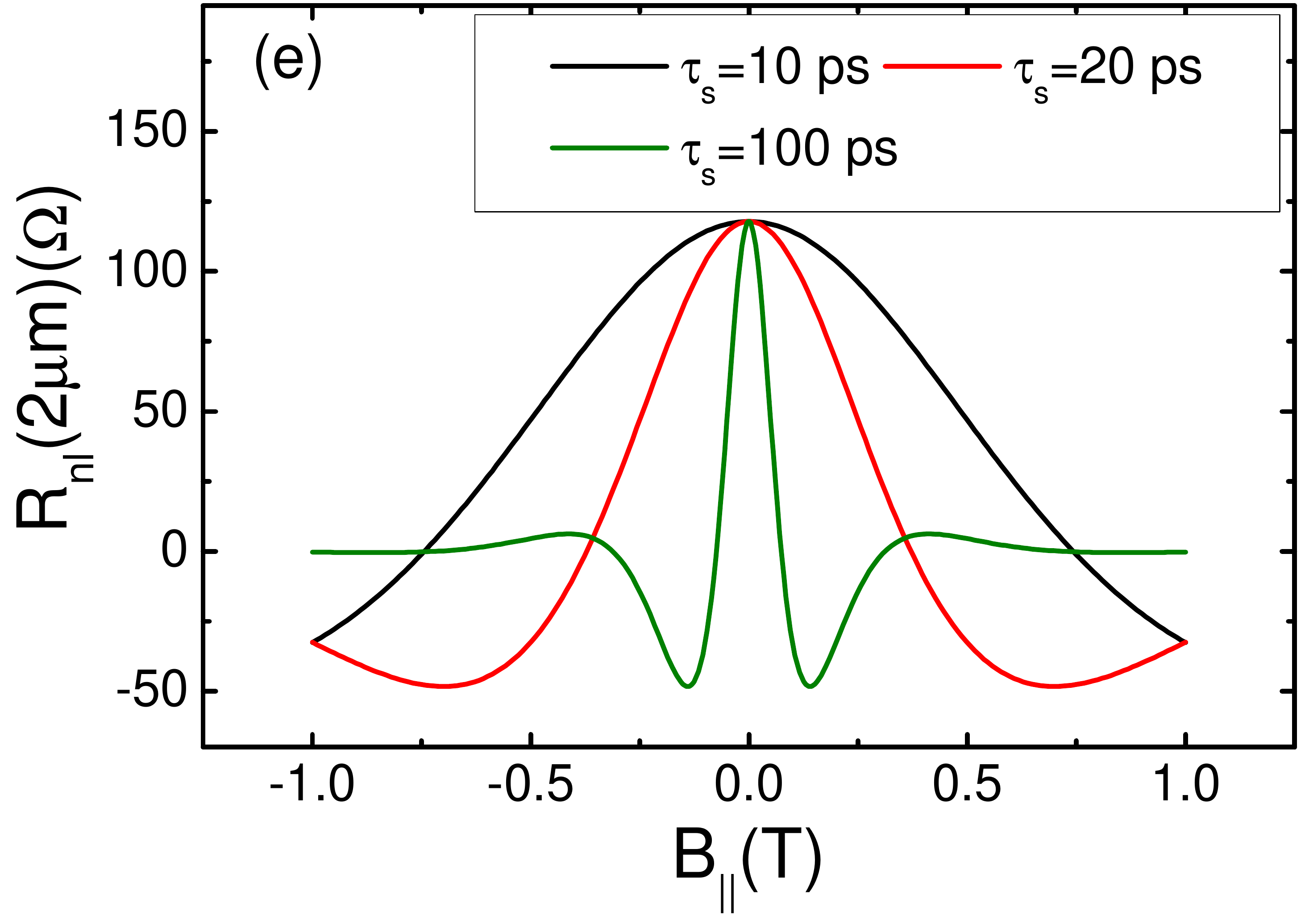}
	\includegraphics[width=0.65\columnwidth]{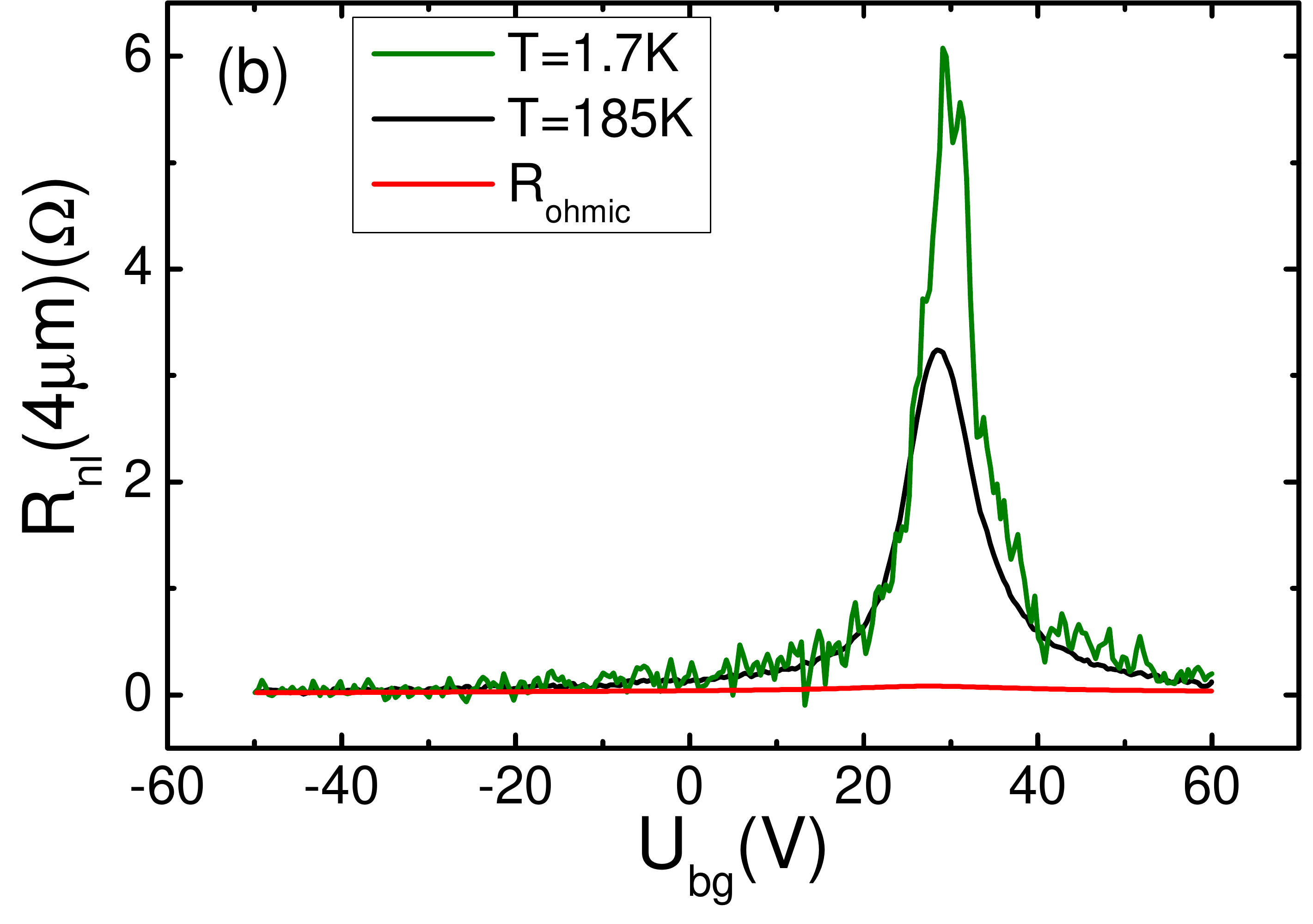}
	\hspace{0.2cm}
	\includegraphics[width=0.65\columnwidth]{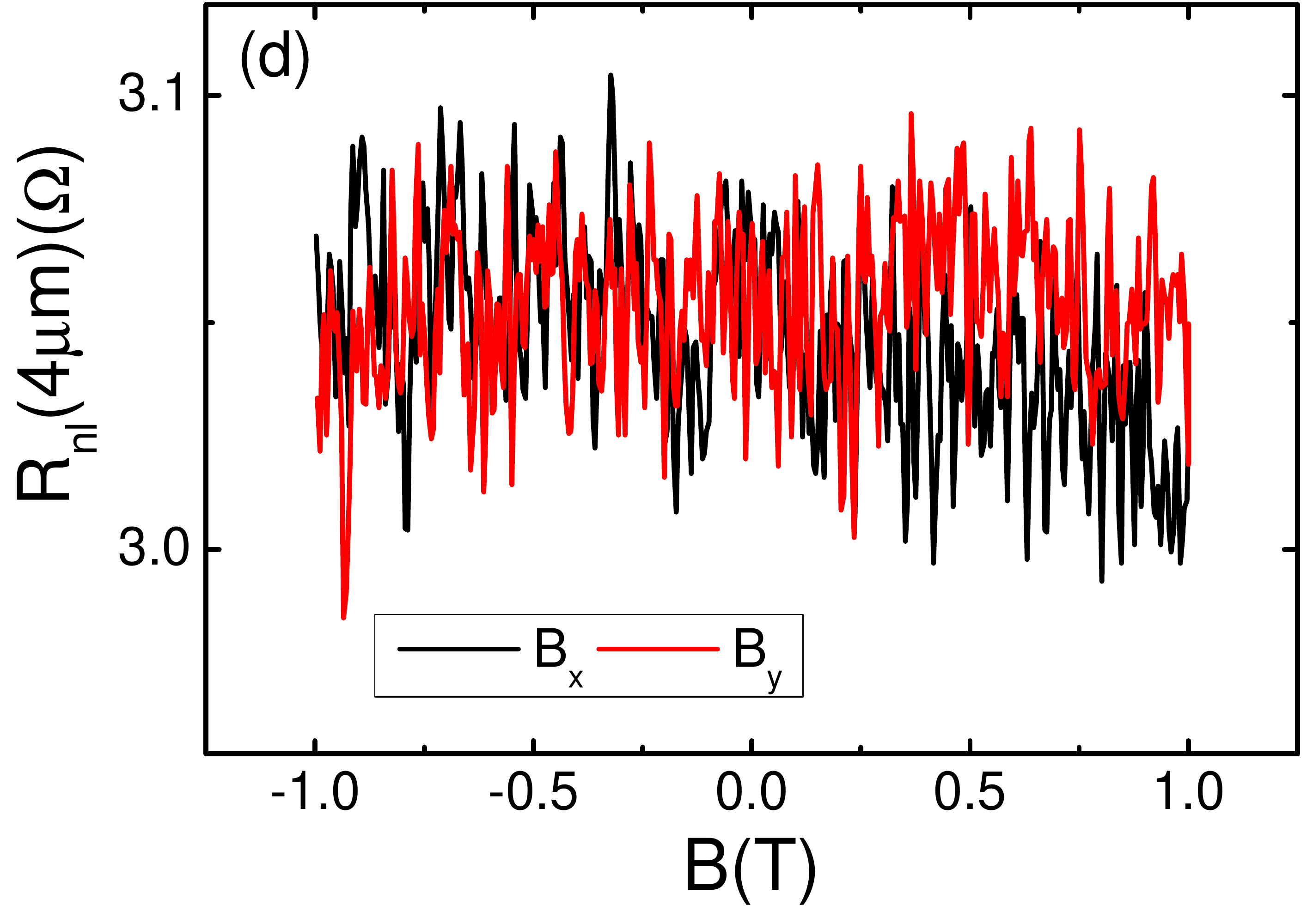}
	\hspace{0.2cm}
	\includegraphics[width=0.65\columnwidth]{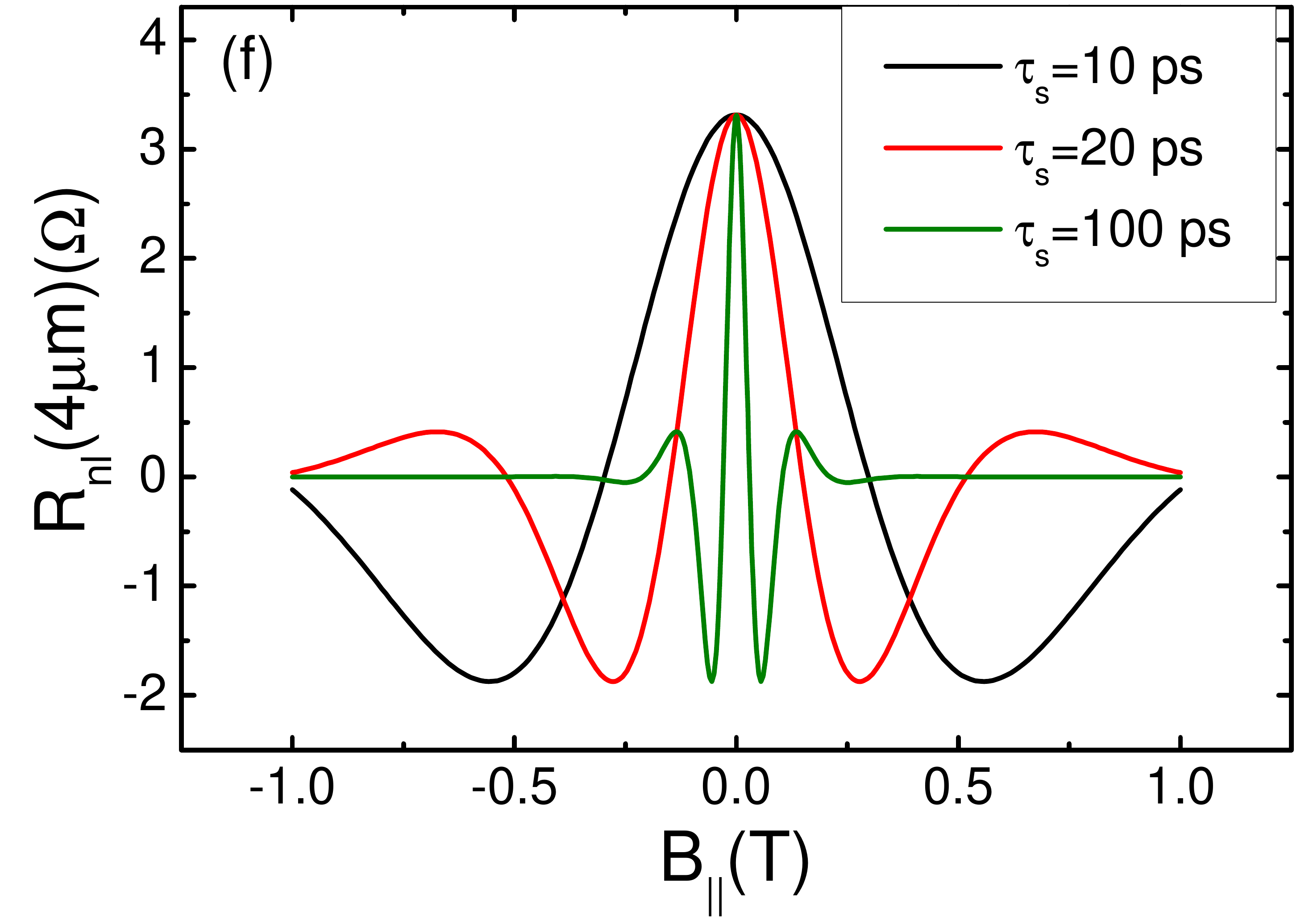}
	\caption[Charge carrier density dependence of nonlocal resistance]{(a) and (b) Charge carrier density dependence of nonlocal resistance measured at two different distances to the current path at $T=185$ K (black curves) and $T=1.7$ K (green curves). In both cases the nonlocal resistance exceeds the expected ohmic contribution (red curves) close to the charge neutrality point. (c) and (d) Dependence of $R_{nl}$ on a magnetic field in both in-plane directions (black and red curves). No noticeable influence of $B_{||}$ on $R_{nl}$ can be observed. (e) and (f) Simulation of the in-plane magnetic field dependence of $R_{nl}$ expected from the spin Hall effect with different spin lifetimes and for different distances from the current path.}
	\label{nl}
\end{figure*}

For obtaining the nonlocal resistance a current was applied between contacts 2 and 8 in the inset of Fig. \ref{hstruct} and a voltage is measured between contacts 3 and 7 (Fig. \ref{nl}(a)) and between contacts 4 and 6 (Fig. \ref{nl}(b)).
Decreasing the temperature from $T=185$ K (black curves in Fig. \ref{nl}(a) and (b)) to $T=1.7$ K (green curves in Fig. \ref{nl}(a) and (b)) increases the nonlocal resistance close to the charge neutrality point. The red curves depict the expected ohmic contribution given by $R_{ohmic}=R_{2pt} \cdot G$, with $R_{2pt}$ being the 2-point resistance between contacts 2 and 8 and a geometry factor $G$ determined by a finite element simulation done with COMSOL. As can be seen in Fig. \ref{nl}(a) and (b), close to the charge neutrality point the measured nonlocal resistances far exceeds the expected ohmic contribution. 

As argued by Balakrishnan~\emph{et al.}\cite{Balakrishnan2013a} this nonlocal resistance might be caused by an interplay between direct and inverse spin Hall effect. Then the nonlocal resistance as a function of distance to the current path $L$ is given by\cite{Abanin2009}:
\begin{equation}
	R_{nl}=\frac{1}{2} \alpha_{SH}^2 \rho \frac{W}{ \lambda_s} \exp \left(-\frac{L}{\lambda_s} \right)
	\label{Rnl}
\end{equation}
with the sheet resistivity $\rho$, the sample width $W$ and the spin diffusion length $\lambda_s$. By comparing $R_{nl}$ at the two different distances in Fig. \ref{nl}(a) and (b) $\lambda_s$ can be calculated to be in the range of $\lambda_s=510 -565$ nm. With this the spin Hall angle $\alpha_{SH}$ close to the charge neutrality point can be calculated to be $\alpha_{SH}=1.3$ for $T=185$ K and $\alpha_{SH}=1.6$ for $T=1.7$ K. These unrealistically high values are similar to the one reported by Kaverzin and van Wees\cite{Kaverzin2015}.

Further, in case that the large nonlocal resistance is caused by the spin Hall effect, $R_{nl}$ should be sensitive to an in-plane magnetic field, due to Larmor precession of the spins. Therefore, an oscillatory behavior of $R_{nl}$ is expected to follow\cite{Abanin2009}:
\begin{equation}
\begin{split}
	R_{nl}(B_{||})= \frac{1}{2} \alpha_{SH}^2 \rho W  
	Re \big[ (\sqrt{1+i \omega_L \tau_s}/\lambda_s)  \\
	\exp (-( \sqrt{1+i \omega_L \tau_s}/\lambda_s) L)  \big]
	\label{nlB}
\end{split}
\end{equation}
with $\omega_L$ being the Larmor frequency.

Fig. \ref{nl}(c) and (d) show the influence of a magnetic field in both in-plane directions (black and red curves) on $R_{nl}$ for two different distances from the current path. As can be seen, no significant change of $R_{nl}$ with $B_{||}$ can be observed. This is in disagreement with the expected behavior given by Eq. \ref{nlB}, which is depicted in Fig. \ref{nl}(e) and (f) for different values of $\tau_s$ in a realistic range, since a lower bound of $\tau_s>10$ ps could be established due to the absence of a weak antilocalization peak\cite{Note1}. As indicated here, a significant dependence of $R_{nl}$ on $B_{||}$ should be visible.

\section{Inverse spin Hall effect in hydrogenated graphene}
Due to the difficulties arising from measuring the spin Hall effect in the H-bar geometry a more direct way for observing this effect is desirable. One way to examine the inverse spin Hall effect electrically was explored by Valenzuela and Tinkham\cite{Valenzuela2006} in aluminum wires. For this they employed electrical spin injection to create a spin current through the wire and measured a resulting nonlocal voltage across a Hall bar.

\begin{figure*}
	\centering
	\includegraphics[width= 0.4\columnwidth]{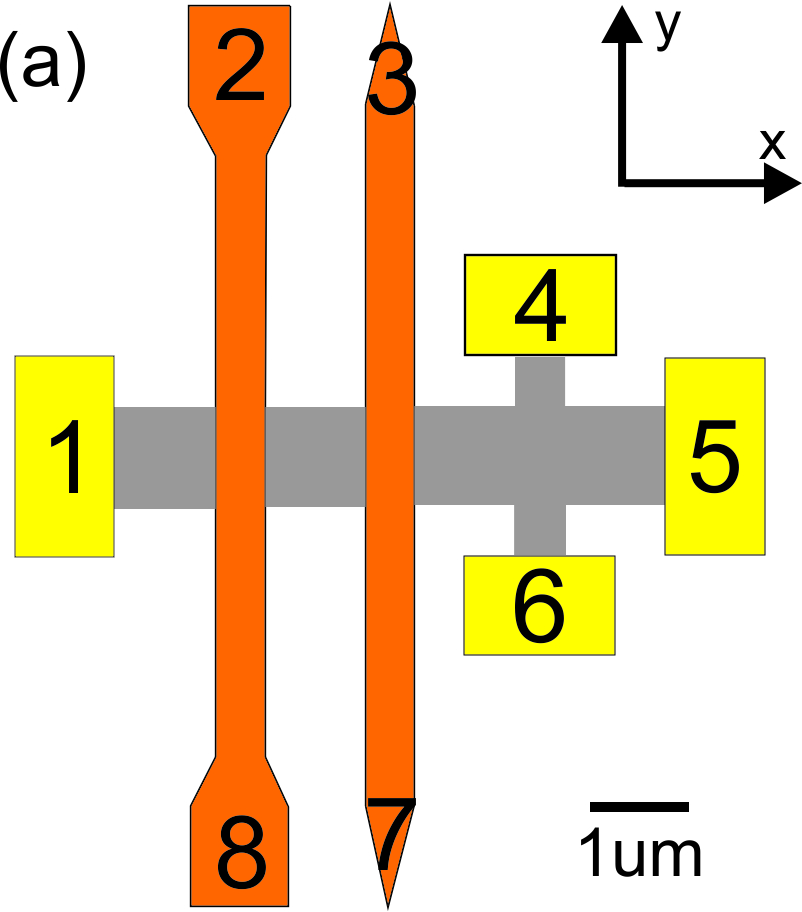}
	\hspace{0.5cm}
	\includegraphics[width=0.7\columnwidth]{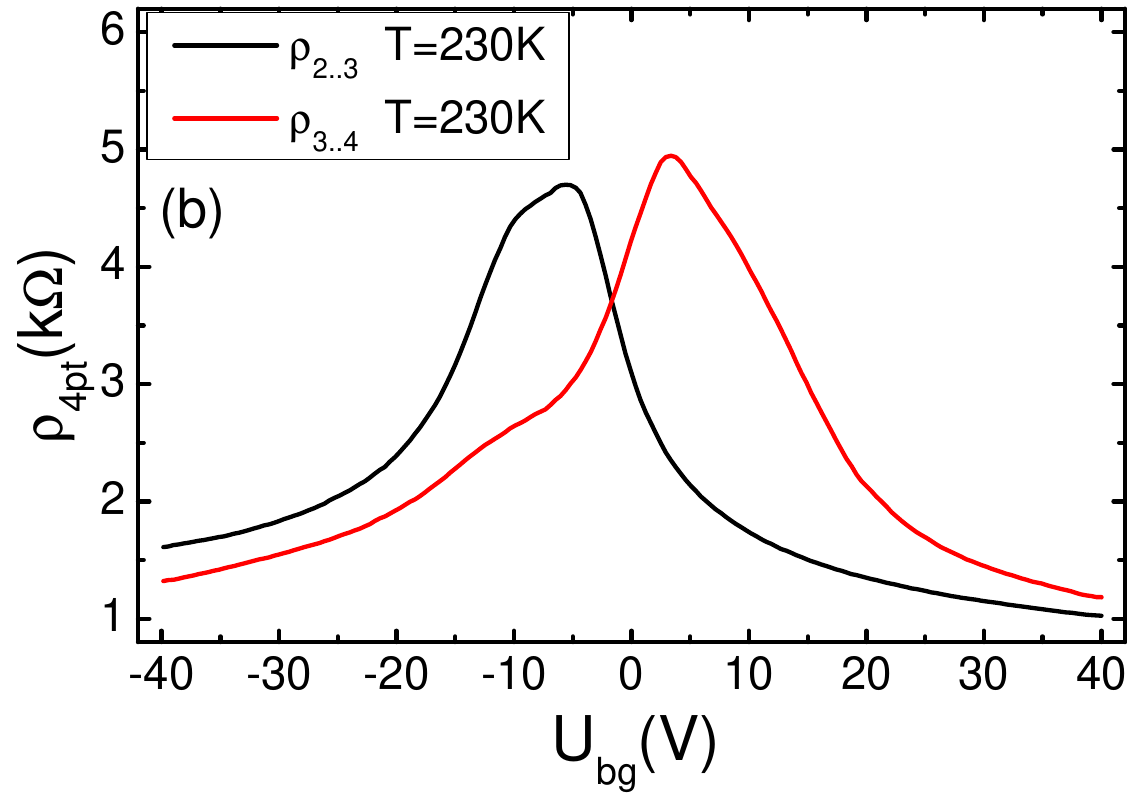}
	\hspace{0.5cm}
	\includegraphics[width=0.7\columnwidth]{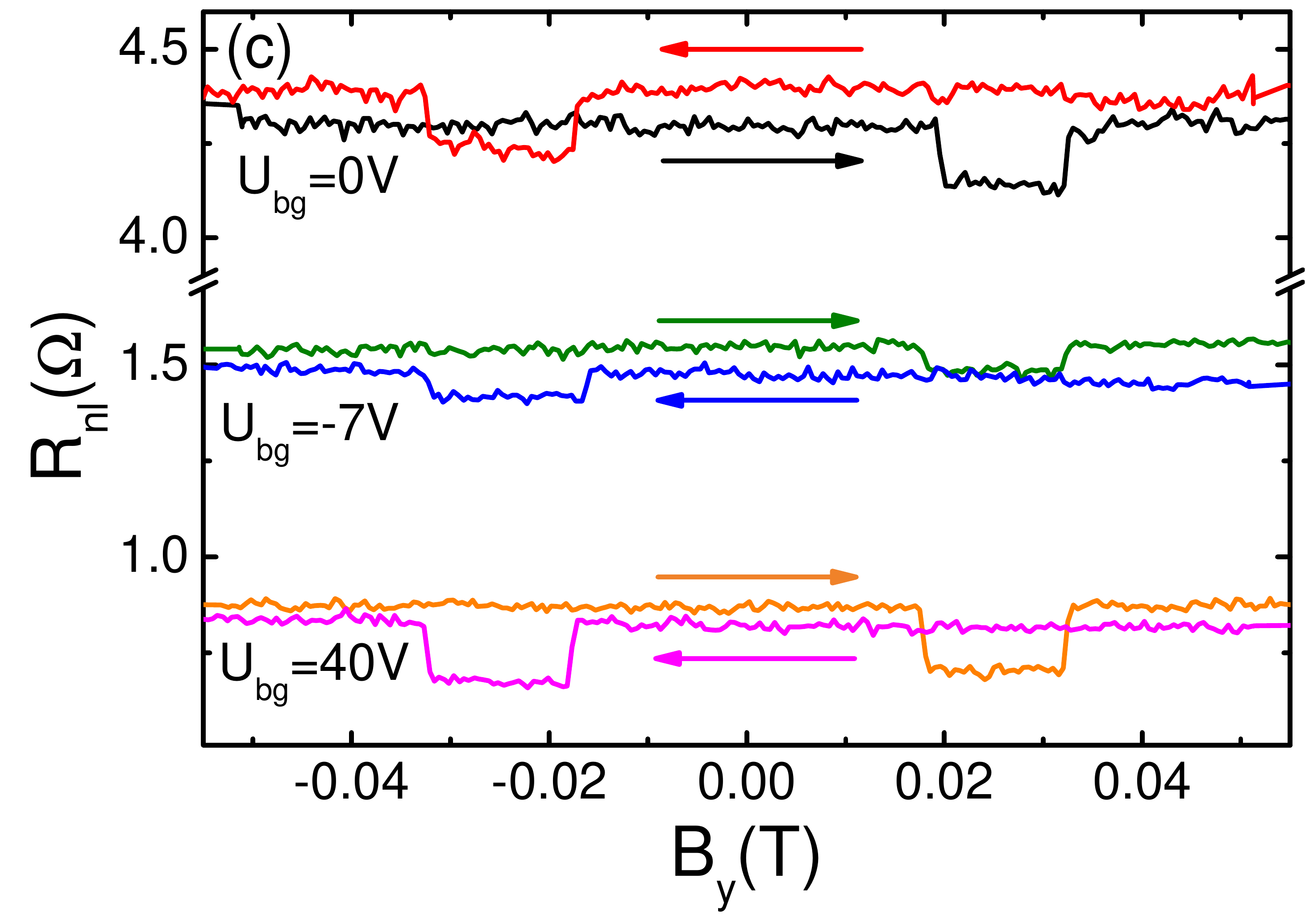}	
	\caption[Sample structure for inverse spin Hall effect in hydrogenated graphene]{(a) Schematic picture of a sample for measuring the inverse spin Hall effect. (b) Back gate sweeps of the inverse spin Hall effect sample. Two areas of the sample (black and red curves) show different doping. (c) Nonlocal spin-valve measurements at different back gate voltages. The reversal of the magnetization of the injection contacts is clearly visible in the nonlocal resistance.}
	\label{06_pic}
\end{figure*}

To employ this method in hydrogenated graphene the sample shown schematically in Fig. \ref{06_pic}(a) was fabricated.
First, exfoliated graphene was exposed to hydrogen plasma for 20 seconds. Spin injection contacts consisting of 1.2 nm MgO, acting as a tunnel barrier, 50 nm Co and 10 nm Au were deposited (orange stripes in Fig. \ref{06_pic}(a)). Afterwards 0.5 nm Cr +80 nm Au were deposited for contacts. As a last step oxygen plasma was employed to etch the sample.

Fig. \ref{06_pic}(b) shows back gate sweeps of this sample, where a current was applied between contacts 1 and 5 and the voltage was taken between contacts 2 and 3 (black curve in Fig. \ref{06_pic}(b)) and between contacts 3 and 4 (red curve in Fig. \ref{06_pic}(b)). As can be seen the position of the charge neutrality point differs for the two areas. This can be caused by different doping of the areas either by the ferromagnetic contacts or by a difference in hydrogen coverage between the area underneath the stripes and the rest of the sample. Mobilities of $\mu_h=2000$ cm$^2$/Vs for the hole side and $\mu_{el}=2400$ cm$^2$/Vs for the electron side could be observed in this sample.

Further, nonlocal spin injection measurements were performed to examine whether spin injection is possible with these contacts\cite{Johnson1985}. Fig. \ref{06_pic}(c) shows nonlocal spin-valve measurements at different back gate voltages. Here a current is applied between contacts 3 and 5 in Fig. \ref{06_pic}(a) and a nonlocal voltage is measured between contacts 2 and 1. The magnetization of the ferromagnetic stripes is first aligned by a magnetic field in stripe direction of $B_y=1$ T. Then the magnetic field is swept in the opposite direction. Due to their different shape the two ferromagnet stripes have a different coercive field. As can be seen in Fig. \ref{06_pic}(c) a clear difference between parallel and antiparallel alignment of the stripe magnetizations can be observed over the whole back gate range.

\begin{figure}
	\centering
	\includegraphics[width=0.9\columnwidth]{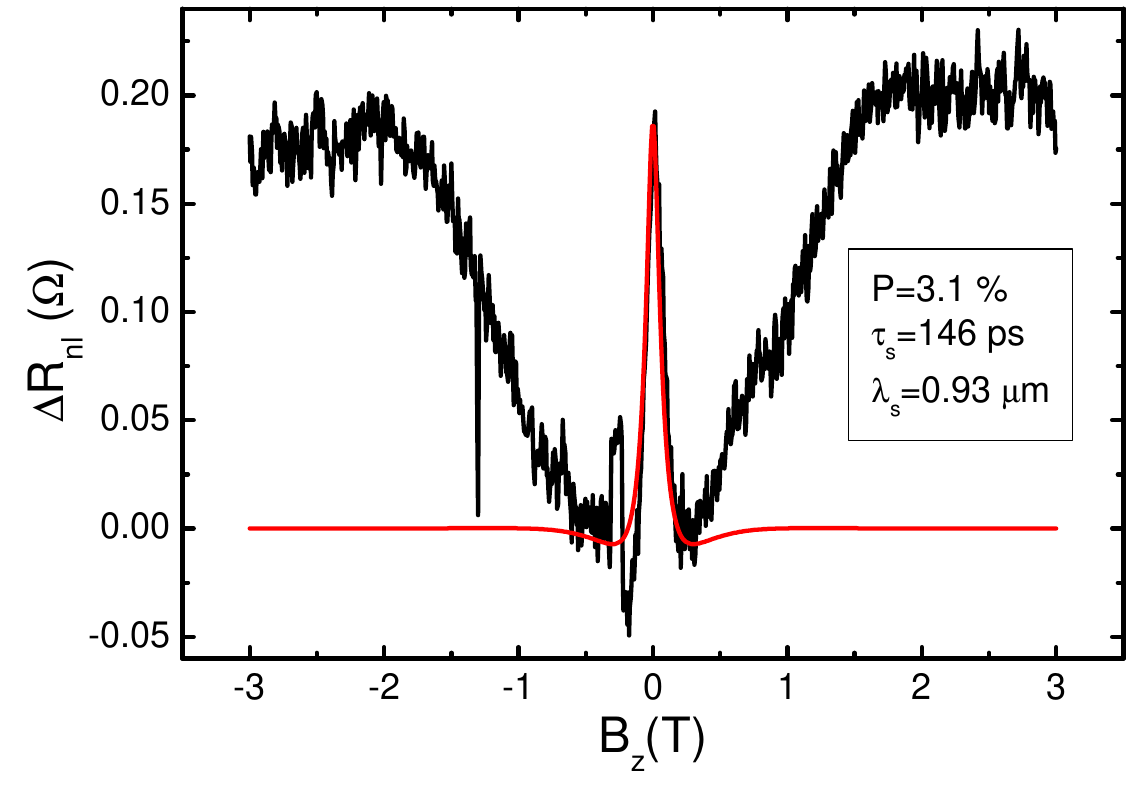}
	\caption[Hanle measurement in hydrogenated field]{Nonlocal resistance after subtraction of a parabolic background (black curve) at $U_{bg}=0$ V. The Hanle peak at low magnetic field can be fitted with Eq. \ref{eqhanle} (red curve). At high magnetic field the stripe magnetization is rotated in the magnetic field direction.}
	\label{06hanle}
\end{figure}

Applying an out-of plane magnetic field to this setup leads to precession of the spins around that field. The out-of plane magnetic field dependence is depicted in Fig. \ref{06hanle}.
Here, a parabolic background that can be caused by a charge current contribution in the nonlocal path by the presence of pinholes in the tunnel barriers\cite{Volmer2015a} was subtracted. In the low magnetic field range the nonlocal resistance follows the expected behavior of the Hanle-effect\cite{Fabian2007}:
\begin{equation}
	\begin{split}
		R_{nl}(\omega_L)=R_{nl}(0) \int_{0}^{\infty} \frac{1}{\sqrt{4 \pi D_s t}} \exp \left( - \frac{L^2}{4 D_s t} \right) \\ \cos(\omega_L t) \exp \left( - \frac{t}{\tau_s} \right) \textrm{d} t\\
		R_{nl}(0)=\frac{P^2 \rho \lambda_s}{2 W} \exp(-L/\lambda_s)
	\end{split}
	\label{eqhanle}
\end{equation}
Fitting the data in the low magnetic field range (red curve in Fig. \ref{06hanle}) reveals a spin injection efficiency of $P=3.1 \%$. The injection efficiency is much lower than what is typically observed with these kind of tunnel barriers in pristine graphene. This can be caused by an enhanced island growth of the MgO tunnel barrier due to the attached hydrogen and therefore an increase of pinholes in the barrier, resulting in a relatively low contact resistance of $R_c=1.2-4.2$ k$\Omega \mu$m$^2$. Another explanation might be increased spin relaxation in the barrier due to the hydrogen atoms. It has to be noted that fabricating spin selective contacts in graphene that was hydrogenated by this method proved to be difficult in general.

Further, the extracted spin lifetime of $\tau_s=146$ ps is much smaller than what was observed in pristine graphene with tunneling contacts produced by the same method\cite{Ringer2018}.This is in contrast to the findings of Wojtaszek~\emph{et al.}~ who observed an increase in spin lifetime after treating pristine graphene with hydrogen plasma\cite{Wojtaszek2013}. This small value for the spin lifetime can be caused by either an increased contact-induced spin relaxation due to an increase in the number of pinholes\cite{Volmer2013} or due to increased spin relaxation by the presence of hydrogen atoms acting as magnetic impurities\cite{Kochan2014}. However, $\tau_s$ is still large enough that a clear oscillation of the nonlocal resistance in the H-bar geometry should be visible as shown by Fig. \ref{nl}(e) and (f).

At higher magnetic fields  the stripe magnetization is rotating into the out-of plane directions. Therefore the polarization of the injected spins has an out-of plane component that does not precess around the external field. The nonlocal resistance saturates around a magnetic field of $B_z=1.8$ T. This value coincides with the field at which the magnetization direction is completely rotated into the out-of plane direction, determined by anisotropic magnetoresistance measurements\cite{Note1}.

Contrary to similar measurements performed by Tombros~\emph{et al.}~ in pristine graphene\cite{Tombros2008a} no difference between the zero magnetic field value and the saturation value of the nonlocal resistance could be observed.
\begin{figure*}
	\centering
	\includegraphics[width=0.64\columnwidth]{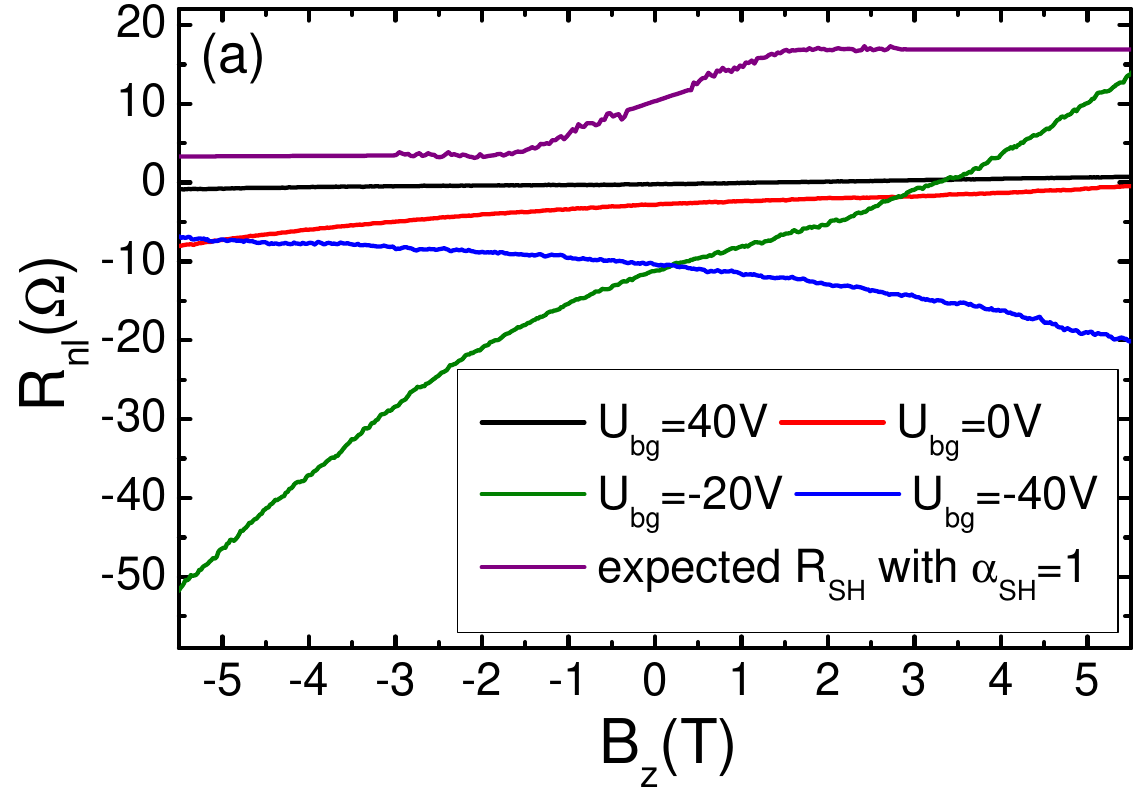}
	\hspace{0.2cm}
	\includegraphics[width=0.64\columnwidth]{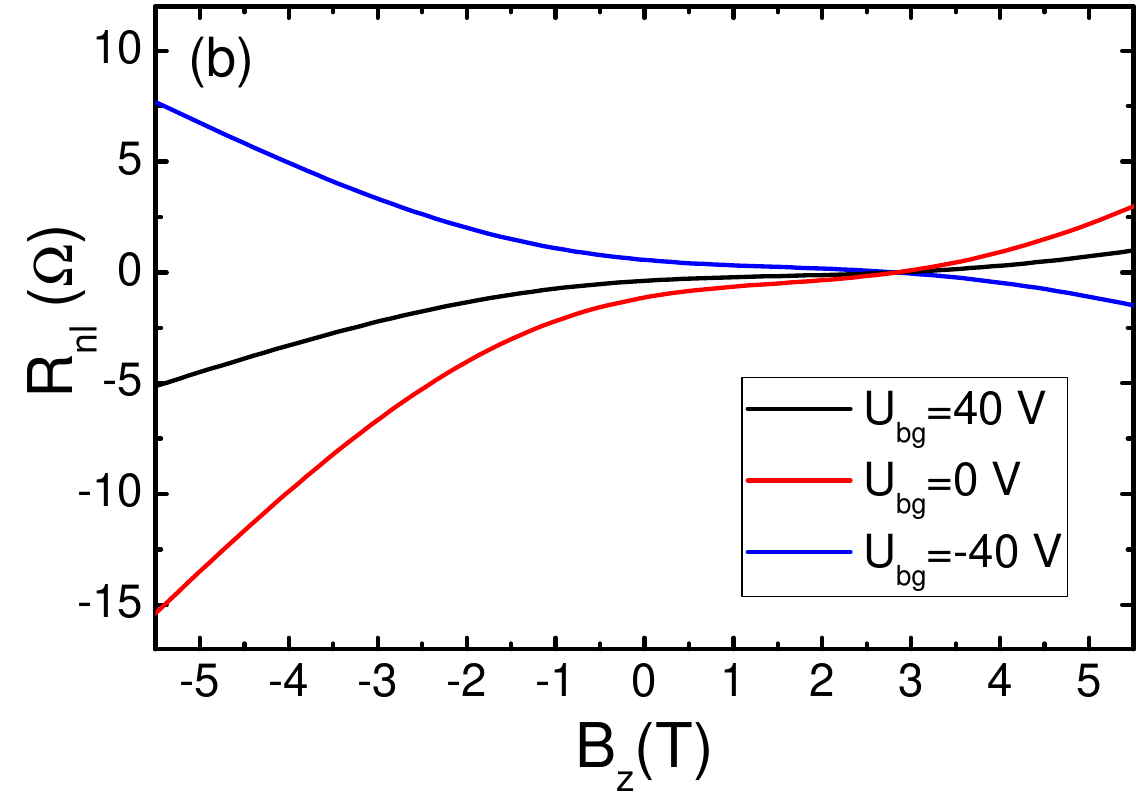}
	\hspace{0.2cm}
	\includegraphics[width=0.64\columnwidth]{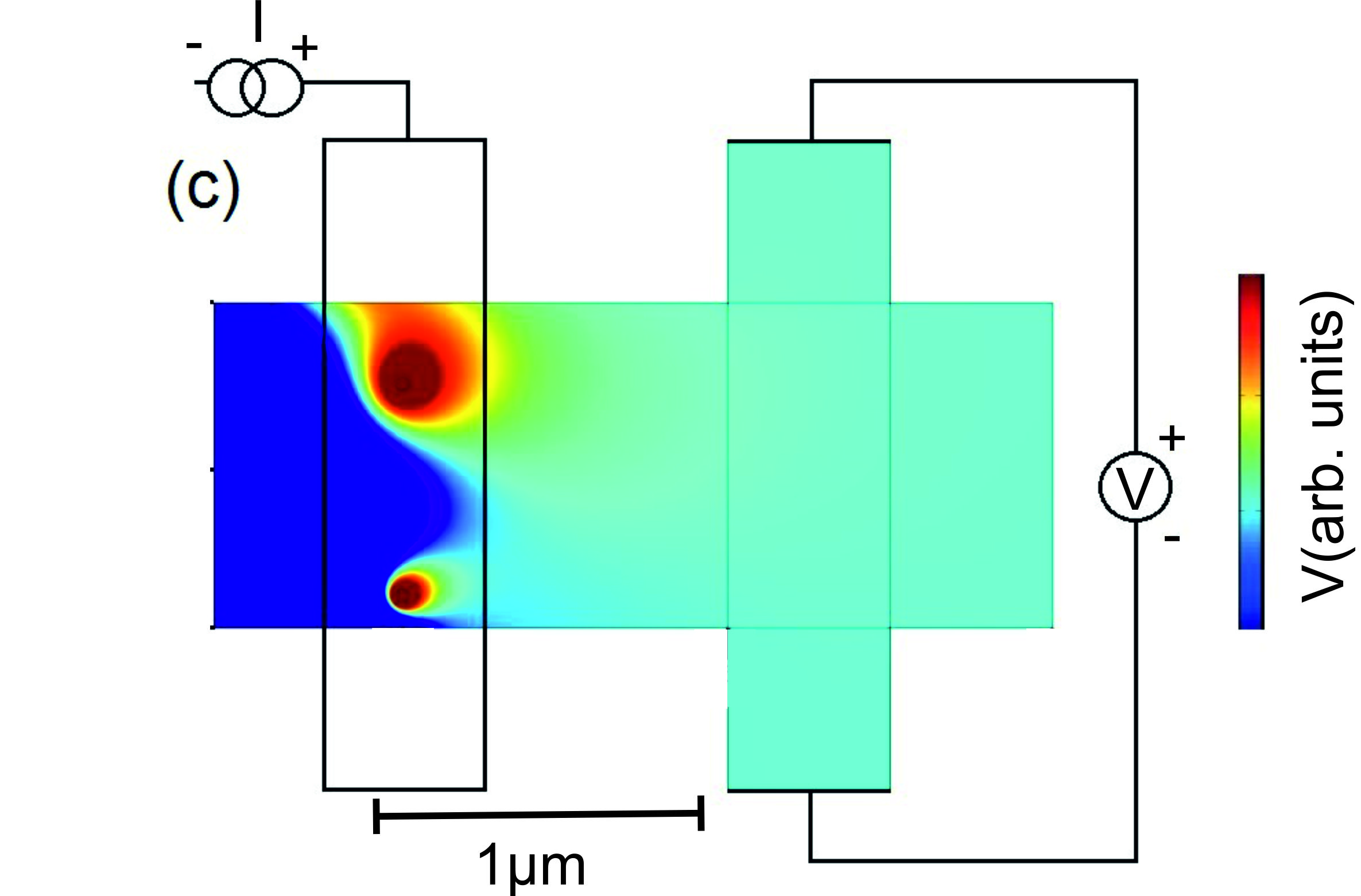}
	\caption[Nonlocal resistance in the inverse spin Hall effect geometry]{(a) Nonlocal resistance in the inverse spin Hall effect geometry at different back gate voltages. No saturation of the nonlocal resistance can be observed. The purple curve depicts the expected $R_{SH}$ given by Eq. \ref{eqishe} with $\alpha_{SH}=1$. (b)  Magnetic field dependent nonlocal resistance for different charge carrier concentration. (c) Potential distribution over the simulated sample in the presence of two pinholes.}
	\label{06ishe}
\end{figure*}
This indicates isotropic spin relaxation, consistent with the expected dominating spin relaxation mechanisms of contact-induced spin relaxation and spin relaxation due to spin-flip scattering at the absorbed hydrogen atoms. Both mechanisms result in isotropic spin relaxation.

For measurement of the inverse spin Hall effect a current was applied between contacts 3 and 1 in Fig. \ref{06_pic}(a) and a nonlocal voltage was measured between contacts 4 and 6. Without an external magnetic field the stripe magnetization is in the in-plane direction. Therefore no nonlocal voltage due to an inverse spin Hall effect is expected. Applying an out-of plane magnetic field results in a rotation of the stripe magnetization towards the out-of plane direction. The resulting out-of plane component of the spin polarization then leads to a nonlocal voltage that is expected to follow\cite{Valenzuela2006}:
\begin{equation}
	R_{SH}=\frac{1}{2} P \alpha_{SH} \rho \exp(-L/\lambda_s) \sin(\theta)
	\label{eqishe}
\end{equation}
with $\sin (\theta)$ being the projection of the stripe magnetization on the $z$-axis. With Eq. \ref{eqhanle} a saturation of the nonlocal resistance at $B_z=1.8$ T with $R_{SH}= \frac{\alpha_{SH} W}{P \lambda_s} R_{nl}(0) \approx \alpha_{SH} \cdot 6.9 \Omega $ is expected. The expected resulting $R_{SH}$ with $\alpha_{SH}=1$ is depicted by the purple curve in Fig. \ref{06ishe}(a).  For this the angular dependence of the magnetization direction $\sin(\theta)$ was extracted from Fig. \ref{06hanle}\cite{Valenzuela2006} and an offset was added for clarity.

The observed nonlocal resistance in this geometry for different back gate voltages is shown in Fig. \ref{06ishe}(a). Here a large magnetic field dependent nonlocal resistance can be seen. However, no saturation of this nonlocal resistance for $B_z>1.8$ T was observed. The magnetic field dependence of the nonlocal resistance is therefore unlikely to be caused by the spin Hall effect.

To determine the origin of this effect a finite element simulation done with COMSOL was performed. For this the potential distribution in the presence of two pinholes in the tunnel barrier was calculated (similar to the calculations in Ref. \onlinecite{Volmer2015a}) as shown in Fig. \ref{06ishe}(c). The resulting magnetic field dependence for different charge carrier concentrations shown in Fig. \ref{06ishe}(b) is comparable to the nonlocal resistance in Fig. \ref{06ishe}(a). Therefore it is likely that the observed magnetic field dependence of the nonlocal resistance is caused by a charge current effect due to the presence of pinholes.

This effect can mask a potential inverse spin Hall effect signal. However, the large spin Hall angle of $\alpha_{SH} \approx 1$ resulting from the spin Hall interpretation of the H-bar geometry should still be observable close to the charge neutrality point $U_{CNP}=10$ V of the areas that are not covered by the ferromagnetic stripes. 

\section{Spin Hall angle - an estimation of order of magnitude}
In this section we provide a theoretical estimate of the upper bound of the spin Hall angle $\alpha_{SH}$ that conventionally expresses a rate conversion of the charge to the transverse spin-current in the presence of SOC. To model hydrogen chemisorption, we employ the tight-binding Hamiltonian inspired by first-principle calculations proposed in Ref. \onlinecite{Gmitra2013}. Plain graphene is described by the conventional nearest-neighbor Hamiltonian $H_0$, and the hydrogen-induced perturbation including a locally enhanced SOC by Hamiltonian $H^\prime$, see Refs. \onlinecite{Gmitra2013, Kochan2017}. Related transport characteristics are estimated on the methodology developed in Refs. \onlinecite{Ferreira2014,Bundesmann2015}. Particularly, for a given scattering process $\mathbf{n}, s\mapsto\tilde{\mathbf{n}}, \tilde{s}$ where an electron with the incident direction and spin, $\mathbf{n}, s$, elastically scatters to an outgoing state $\tilde{\mathbf{n}},\tilde{s}$, we calculate the corresponding differential cross-section $\tfrac{\mathrm{d}\sigma}{\mathrm{d}\varphi}\bigl(\mathbf{n},s;\tilde{\mathbf{n}},\tilde{s}\bigr)$ that depends also on the energy of the incident electron. Knowing $\tfrac{\mathrm{d}\sigma}{\mathrm{d}\varphi}$ we know spatial probability distributions of electrons with flipped or conserved spin depending on the relative angle $\tilde{\varphi}_{\mathbf{n}}=\sphericalangle(\mathbf{n}\tilde{\mathbf{n}})$.
Elastic scattering governed by $H^\prime$ affects momentum relaxation due to resonances near the Dirac point\cite{Wehling2010, Irmer2018}, and also spin relaxation due to locally enhanced SOC\cite{Bundesmann2015}. Despite the fact that hydrogen is predicted to induce also an unpaired magnetic moment\cite{Yazyev2007}, which can serve as another spin relaxation channel\cite{Kochan2014}, we restrict our estimates of $\alpha_{SH}$ just to the local SOC interactions.

\begin{figure}
	\includegraphics[width=0.45\textwidth]{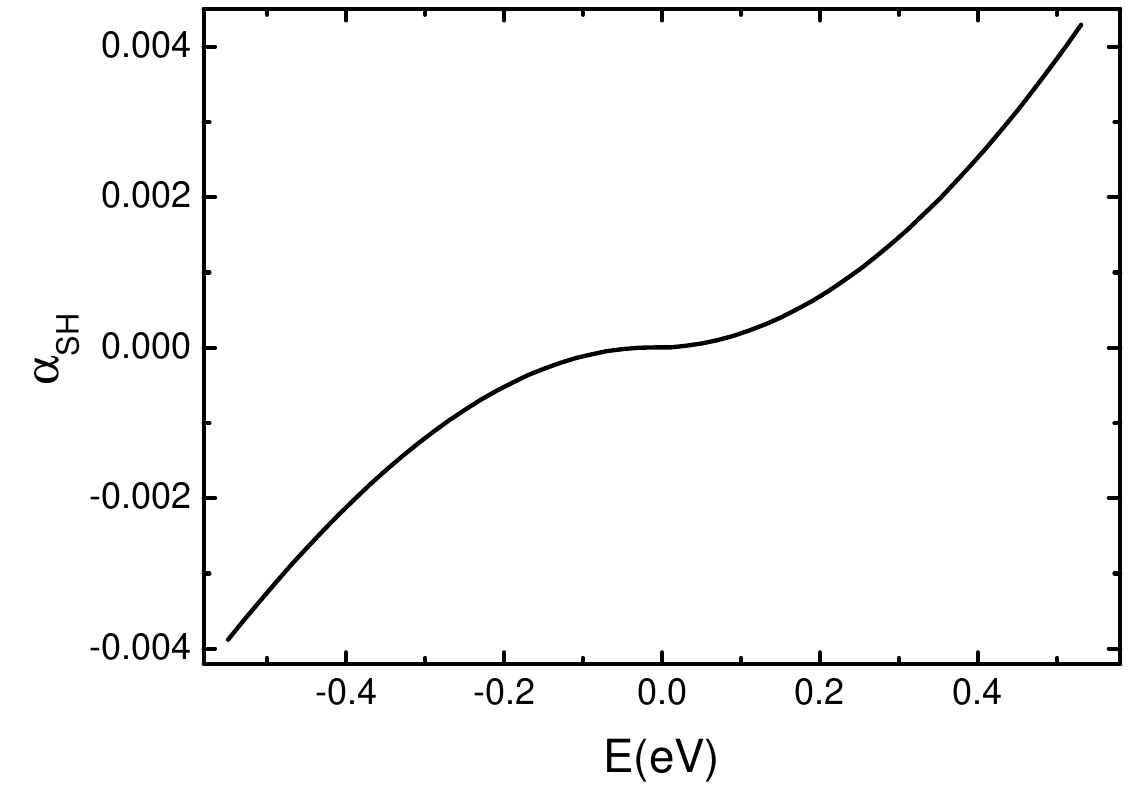}
	\caption{Estimated spin Hall angle $\alpha_{SH}$ at zero temperature for a dilute hydrogenated graphene as a function of Fermi energy (zero energy corresponds to the charge neutrality point). The tight-binding parameters and model-based calculation follow\cite{Gmitra2013,Bundesmann2015}.\label{FigX}}
\end{figure}

Assuming a spin polarized beam of, say, spin-up electrons with the incident energy $E$, the upper bound of the spin Hall angle $\alpha_{SH}(E)$ reads:
\begin{equation}
\alpha_{SH}(E)\approx \frac{\left\langle\sum\limits_{\tilde{\mathbf{n}}}\left[\frac{\mathrm{d}\sigma}{\mathrm{d}\varphi}\bigl(\mathbf{n},\uparrow;\tilde{\mathbf{n}},\uparrow\bigr)-
	\frac{\mathrm{d}\sigma}{\mathrm{d}\varphi}\bigl(\mathbf{n},\downarrow;\tilde{\mathbf{n}},\uparrow\bigr)\right]\sin{\tilde{\varphi}_{\mathbf{n}}}\right\rangle}
{\left\langle\sum\limits_{\tilde{\mathbf{n}}}\left[\frac{\mathrm{d}\sigma}{\mathrm{d}\varphi}\bigl(\mathbf{n},\uparrow;\tilde{\mathbf{n}},\downarrow\bigr)\right]
	2\cos{\tilde{\varphi}_{\mathbf{n}}}\right\rangle}\,,
\end{equation}
where the angle brackets represent averaging over all incoming directions $\mathbf{n}$. The calculation was performed for one hydrogen atom in a supercell containing 16120 carbon atoms, {\em i.e.} a hydrogen concentration of 0.0062 \%. 
Fig. \ref{FigX} displays $\alpha_{SH}$ as function of Fermi energy. The obtained values are in magnitude comparable with, {\em e.g.}, those of Ferreira~\emph{et al.}~\cite{Ferreira2014}, but differ from the experimental data fitted by Eq. \ref{Rnl}. 
Further, as seen in Fig. \ref{FigX}, $\alpha_{SH}$ is expected to vanish at the charge neutrality point, which is in contrast to the observed nonlocal resistance in Fig. \ref{nl}.

\section{Discussion}
The background effect observed in Fig. \ref{06ishe} could mask the relatively small spin Hall angle resulting from the theoretical estimation in Fig. \ref{FigX}. However, the high value of $\alpha_{SH}>1$ following from the SHE interpretation of the nonlocal resistance in Fig. \ref{nl}(a) and (b) should still be observable. Further, this unusually high spin Hall angle as well as the absence of an oscillatory behavior of $R_{nl}$ with an in-plane magnetic field support the findings of Kaverzin and van Wees\cite{Kaverzin2015}. These results suggest that the large nonlocal resistance observed in Fig. \ref{nl}(a) and (b) is caused by a non spin-related mechanism.

Large nonlocal resistances in the H-bar structure were also observed in graphene decorated with heavy atoms\cite{Wang2015}, hBN/graphene heterostructures\cite{Gorbachev2014} and in graphene structured with an antidot array\cite{Pan2017}. These were attributed to the occurrence of a valley-Hall effect\cite{Wang2015,Gorbachev2014}, a nonzero Berry curvature, due to the presence of a band gap\cite{Pan2017} and transport through evanescent waves\cite{VanTuan2016,Tworzydlo2006}. However none of these effects can sufficiently explain the observed behavior\cite{Note1}.

\section{Conclusion}
In conclusion we employed two different types of measurements to investigate the spin Hall effect in hydrogenated graphene. For hydrogenation, graphene was placed into a hydrogen plasma. This technique was investigated by Raman spectroscopy. Since Raman measurements are only sensitive to the number of defects and not to the defect type, measurements with both hydrogen and deuterium were performed. The different desorption behavior observed for these isotopes is a clear indication that the defects produced by this method are indeed bonded hydrogen atoms.

Nonlocal measurements in the so called H-bar geometry showed a large nonlocal resistance that however did not show a dependence on an in-plane magnetic field. Also measurement of the inverse spin Hall effect by electrical spin injection showed no sign of the large spin Hall angle suggested by the spin Hall effect interpretation of the nonlocal measurements. Further, a theoretical estimate showed a much smaller spin Hall angle than suggested by the spin Hall interpretation of the nonlocal resistance in the H-bar method. These results indicate that the large nonlocal resistance is caused by a non spin-related origin. 

\section*{Acknowledgments}
Financial support by the Deutsche Forschungsgemeinschaft (DFG) through project KO 3612/3-1 and within the programs GRK 1570, SFB 689, and SFB 1277 (projects A09, B05 and B06) is gratefully acknowledged. This project has received funding from the European Union's Horizon 2020 research and innovation program under grant agreement No 696656 (Graphene Flagship).



\section*{Supplementary material}

# Absence of a giant spin Hall effect in plasma-hydrogenated graphene
## (Supplemental Material)

## I. GRAPHENE HYDROGENATED WITH HSQ

For comparison samples were produced for which hydrogenation was done by exposing hydrogen silsesquioxane (HSQ) with an electron beam. Fig. S1 (a) shows the Raman spectrum of such a sample with D- to G-peak intensity ratio of: $I_D/I_G = 0.56$, resulting in a calculated hydrogen coverage of 0.0033 % that is comparable to the sample described in the main text. As can be seen in Fig. S1 (b) this method for hydrogenation resulted in a very high doping of the sample with $U_{CNP} > 100$ V. Therefore the charge neutrality point could not be reached and no nonlocal resistance in the H-bar method could be observed.

## II. WEAK LOCALIZATION AND ANTILOCALIZATION IN PLASMA HYDROGENATED GRAPHENE

A possible signature of strong spin-orbit coupling in a material is the occurrence of a weak antilocalization effect. Recently pronounced weak antilocalization caused by spin-orbit coupling was reported in graphene/transition metal dichalcogenide heterostructures [1–4]. From these measurements a spin-orbit scattering time, which is an upper bound to the spin lifetime, of $\tau_{so} = 0.2 - 5$ ps could be extracted.

Magnetoresistance measurements of the sample shown in Fig. 2 (main text) were performed in a 4-point configuration at temperature $T = 1.7$ K. Fig. S2 (a) shows the out-of plane magnetic field dependence of the conductivity at different gate voltages. For these a parabolic background (due to positive magnetoresistance) was subtracted. Further, in order to suppress universal conductance fluctuations, for the curve close to the charge neutrality point, an average over 14 curves at slightly different back gate voltages was taken. The sharp dip in the magnetoconductance indicates a weak localization effect. For single layer graphene weak localization can be described by [5,6]:

$$\Delta\sigma_{WL} = \frac{e^2}{\pi h}\left[ F\left(\frac{\tau_B^{-1}}{\tau_\phi^{-1}}\right) - F\left(\frac{\tau_B^{-1}}{\tau_\phi^{-1} + 2\tau_i^{-1}}\right) - 2F\left(\frac{\tau_B^{-1}}{\tau_\phi^{-1} + \tau_i^{-1} + \tau_*^{-1}}\right) \right] \tag{S1}$$

with $F(z) = \ln(z) + \Psi(1/2 + 1/z)$, the Digamma function $\Psi(x)$ and $\tau_B^{-1} = 4eD_cB/\hbar$. Here $\tau_\phi$ denotes the phase coherence time, $\tau_i$ the intervalley scattering time and $\tau_*$ the intravalley scattering time. Fitting the magneto-conductivity correction with this formula (red curves in Fig. S2 (a)) gives values of $\tau_\phi = 7.0$ ps ($\tau_\phi = 9.2$ ps) and

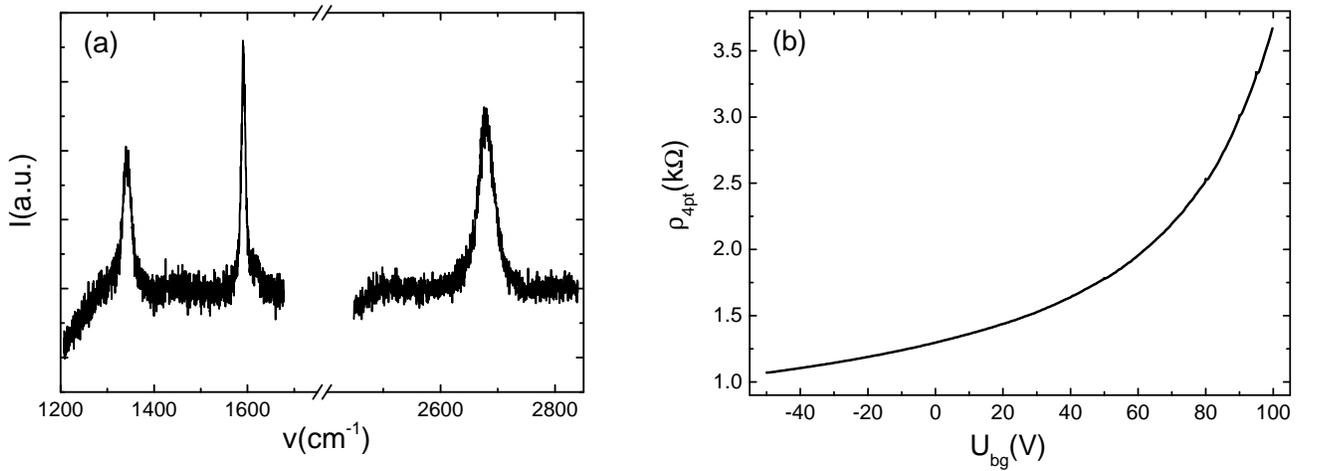

FIG. S1. (a) Raman spectrum of a sample for which hydrogenation was done by exposing hydrogen silsesquioxane (HSQ) with an electron beam. The ratio between D- and G-peak intensities of $I_D/I_G = 0.56$ indicates a hydrogen coverage of 0.0033 %. (b) Back gate dependent four-point resistivity of this sample. The high doping prevents measurements close to the charge neutrality point.

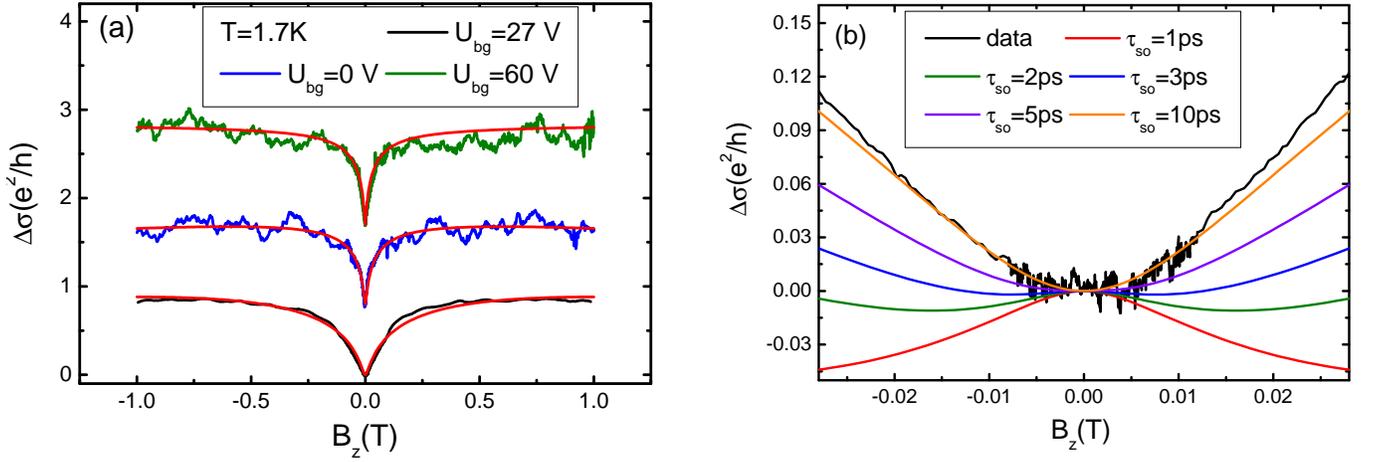

FIG. S2. (a) Magnetoconductivity in hydrogenated graphene at different back gate voltages. The peaks around $B_z = 0\,\mathrm{T}$ can be fitted with Eq. S1 indicating a weak localization effect. (b) Low magnetic field correction to the magnetoconductivity according to Eq. S2 at $U_{bg} = 27\,\mathrm{V}$. The absence of a weak antilocalization peak gives a lower bound for $\tau_{so}$.

$\tau_i = 0.68\,\mathrm{ps}\,(\tau_i = 0.53\,\mathrm{ps})$ for $U_g - U_{CNP} = -30\,\mathrm{V}\,(U_g - U_{CNP} = 30\,\mathrm{V})$. The magnetoconductivity close to the charge neutrality point $U_{CNP} = 27\,\mathrm{V}$, shown by the black curve in Fig. S2 (a), gives the values $\tau_\phi = 2.6\,\mathrm{ps}$ and $\tau_i = 0.077\,\mathrm{ps}$. The lower value of $\tau_\phi$ close to the charge neutrality point as compared to $U_g - U_{CNP} = \pm 30\,\mathrm{V}$ is common in graphene and can be explained by increased electron-electron interaction at lower charge carrier concentration [7]. However the intervalley scattering time $\tau_i$ was observed to be independent on charge carrier density in pristine graphene [7]. The large charge carrier density dependence of $\tau_i$ could therefore be a indication of resonant scattering by the hydrogen atoms [8,9].

As can be seen by Eq. S1 a positive value for the conductivity correction and therefore weak antilocalization can be achieved for small ratios of $\tau_\phi/\tau_i$ and $\tau_\phi/\tau_*$. This is usually realized by increasing the temperature which decreases $\tau_\phi$ [10]. However, at low temperatures weak antilocalization can be observed in graphene when strong spin-orbit coupling is present. For this case the corresponding conductivity correction was calculated by McCann and Falko to be [11]:

$$\Delta\sigma_{WAL} = -\frac{e^2}{2\pi h}\left[F\left(\frac{\tau_B^{-1}}{\tau_\phi^{-1}}\right) - F\left(\frac{\tau_B^{-1}}{\tau_\phi^{-1} + 2\tau_{asy}^{-1}}\right) - 2F\left(\frac{\tau_B^{-1}}{\tau_\phi^{-1} + \tau_{so}^{-1}}\right)\right] \tag{S2}$$

where the spin-orbit scattering time $\tau_{so}$ combines the contributions of spin-orbit coupling that are symmetric ($\tau_{sym}$) and asymmetric ($\tau_{asy}$) in the $z \to -z$ direction: $\tau_{so}^{-1} = \tau_{sym}^{-1} + \tau_{asy}^{-1}$. This formula is only valid for the case of strong intervalley scattering $\tau_i^{-1} > \tau_\phi^{-1}, \tau_{so}^{-1}, \tau_{asy}^{-1}$ and low magnetic field. It has to be noted that in the absence of asymmetric spin-orbit coupling, $\tau_{sym}$ only leads to a suppression of weak localization and no weak antilocalization peak is expected [11].

Since the spin-Hall interpretation of the nonlocal resistance in the H-bar method suggests an increased SOC close to the charge neutrality point, the low magnetic field range in this back gate range was examined. The black curve in Fig. S2 (b) shows the low magnetic field behavior at $U_{bg} = 27\,\mathrm{V}$. As can be seen, no weak antilocalization peak could be observed for this sample. Further, Fig. S2 (b) shows from Eq. S2 predicted curves with several values of $\tau_{so}$. Here the spin-orbit coupling was assumed to be completely asymmetric ($\tau_{so} = \tau_{asy}$) and the phase coherence time $\tau_\phi = 2.6\,\mathrm{ps}$ from the weak localization fit was taken. A clear weak localization is only expected for $\tau_{so} < \tau_\phi$. Due to the relatively low phase coherence time in this sample compared to the results from the graphene/TMDC heterostructures only a lower bound of $\tau_{asy} \geq 10\,\mathrm{ps}$ can be established from this.

## III. ANISOTROPIC MAGNETORESISTANCE MEASUREMENTS

The dependence of the injector stripe magnetization on an external magnetic field can be probed by anisotropic magnetoresistance (AMR) measurements. Here the stripe resistance follows the relation [12]:

$$R(\phi) = R_0 + \Delta R \cos^2(\phi) \tag{S3}$$

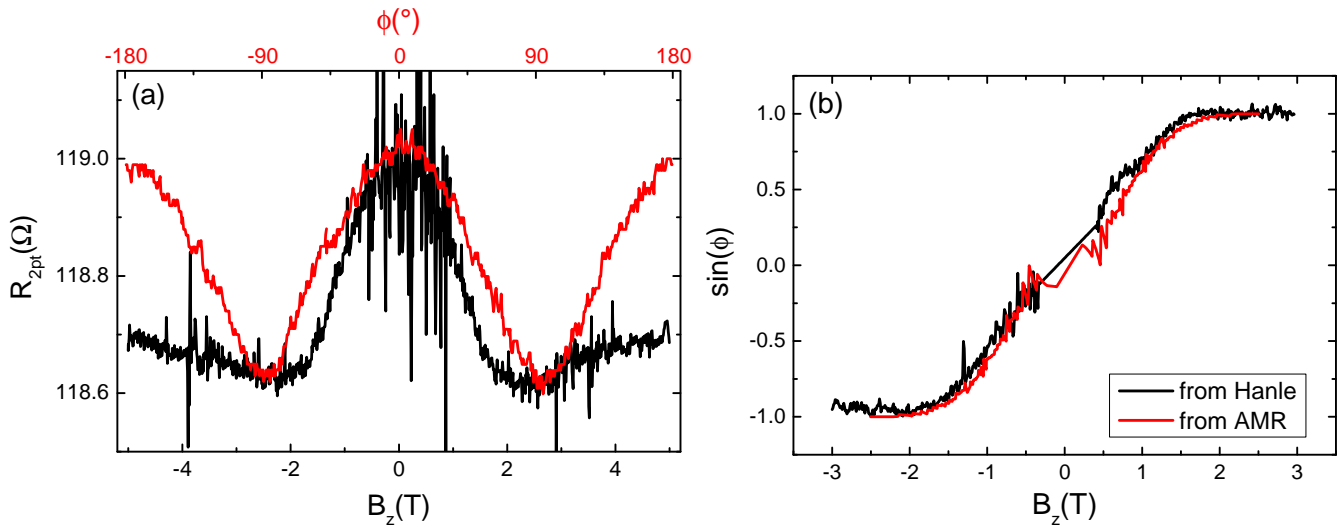

FIG. S3. (a) Anisotropic magnetoresistance of the ferromagnetic injector stripe with regards to an out-of plane magnetic field (black curve) and an inplane magnetic field of $B = 1$ T at varying angle between magnetic field and stripe orientation $\phi$. (b) Projection of the stripe magnetization on the z-axis determined by Hanle-measurements (black curve) and anisotropic magnetoresistance measurements (red curve).

with $\phi$ being the angle between magnetization and current direction. Therefore the stripe resistance has a maximum for parallel current and magnetization directions and a minimum for perpendicular current and magnetization directions.

The red curve in Fig. S3 (a) shows the two-point resistance of the ferromagnetic stripe (between contacts 3 and 7 in Fig. 4 (a)) under rotation of an inplane magnetic field of $B = 1$ T. As can be seen the stripe resistance follows the $\cos^2$ relation from Eq. S3 with regards to the angle between stripe (and current) and magnetic field directions.

The resistance of the ferromagnetic stripe in dependence of an out-of plane magnetic field is shown by the black curve in Fig. S3 (a). For this a linear background was subtracted. As can be seen the stripe resistance decreases with magnetic field as the magnetization rotates towards the out-of plane direction. A saturation is reached when the magnetization is completely rotated to the out-of plane direction at a magnetic field of $B = 1.8$ T.

By comparing the black and red curve in Fig. S3 (a) the projection of the stripe magnetization can be calculated (red curve in Fig. S3 (b)). As can be seen this yields the same results as extracting this behavior from the Hanle measurements from Fig. 5 which was used to determine the expected nonlocal resistance from the spin-Hall effect in Fig. 6 (a).

## IV. DISCUSSION OF ORIGIN OF THE NONLOCAL RESISTANCE IN THE H-BAR METHOD

The absence of both a weak antilocalization effect and an inverse spin-Hall effect in hydrogenated graphene suggests that the large nonlocal resistance observed in Fig. 3 (a) and (b) is caused by a non spin-related mechanism. A similar behavior (large nonlocal resistance without any dependence on an inplane magnetic field) was also observed by Wang et al. in graphene decorated with heavy atoms[13] with a similar relaxation length $\lambda = 250 - 400$ nm. They attribute this effect to a valley-Hall effect. This effect was observed in hBN/graphene heterostructures, where alignment between hBN and graphene can break the symmetry between the graphene sublattices and therefore produces a valley polarized current[14]. However, for hydrogenated graphene the occurrence of a sublattice asymmetry is unlikely as this would mean a preference in the hydrogen binding to one of the sublattices.

Further, large nonlocal resistances were observed by Pan *et al.* in graphene structured with an antidot array[15]. They argue that even in the absence of sublattice asymmetry the presence of a band gap is sufficient to introduce a nonzero Berry curvature, that in turn produces the nonlocal resistance. From the temperature dependence of the resistivity at the charge neutrality point for this sample (not shown here) a band gap of $E_g = 0.25$ meV could be extracted. Pan et al. calculated the expected nonlocal resistance to be[15]:

$$R_{nl} = \rho^3 \left(\frac{e^2}{h}\right)^2 \frac{(E_g/\hbar v)^2}{(E_g/\hbar v)^2 + \pi\sqrt{n^2 + n_0^2}} \cdot a \tag{S4}$$

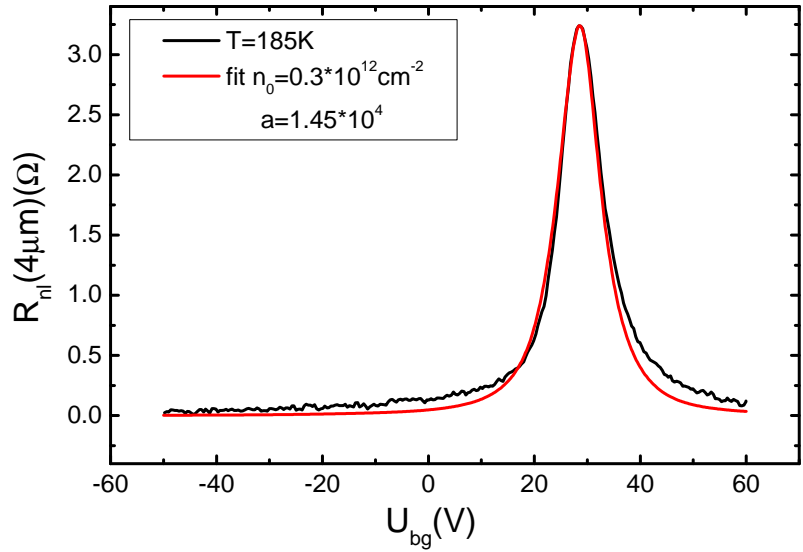

FIG. S4. Fit of back gate dependent nonlocal resistance with Eq. S4. The required coefficient $a = 1.45 \cdot 10^4$ is much too large to explain the measured nonlocal resistance.

with the Fermi velocity $v$, the charge carrier concentration $n$, the residue charge carrier concentration at the charge neutrality point $n_0$ and $a$ a coefficient that is dependent on the sample geometry and the exponential decay length. Fig. S4 depicts a fit of the nonlocal resistance with Eq. S4. As can be seen a good agreement between measurement and theoretical prediction can be achieved with the coefficient $a = 1.45 \cdot 10^4$. However, $a$ is expected to be of the order $a \approx e^{-L/\lambda} = 6.9 \cdot 10^{-4}$. Therefore the expected magnitude of this effect is much too small to explain the measured nonlocal resistance.

Finally, calculations by Van Tuan et al. [16] showed that for gold decorated graphene a sizeable contribution to the nonlocal resistance is generated by transport through evanescent waves [17]. However, this effect is expected to dominate for $W > L$, which is contrary to our observed nonlocal resistance at with $L = 2W$ (Fig. 3 (a)) and $L = 4W$ (Fig. 3 (b)).



\end{document}